\documentclass[11pt]{article}
\pdfoutput=1
\usepackage{amsmath}
\usepackage{amsthm}
\usepackage{amscd}
\usepackage{amsfonts}
\usepackage[psamsfonts]{amssymb}
\usepackage{graphicx}
\usepackage{hyperref}
\usepackage{xcolor}
\usepackage{bm}

\makeatletter
\renewcommand\section{\@startsection {section}{1}{\z@}%
                                 {-3.5ex \@plus -1ex \@minus -.2ex}%nn
                                   {2.3ex \@plus.2ex}%
                                   {\normalfont\large\bfseries}}
\renewcommand\subsection{\@startsection{subsection}{2}{\z@}%
                                   {-3.25ex\@plus -1ex \@minus -.2ex}%
                                     {1.5ex \@plus .2ex}%
                                     {\normalfont\bfseries}}
\renewcommand\subsubsection{\@startsection{subsubsection}{3}{\z@}%
                                   {-3.25ex\@plus -1ex \@minus -.2ex}%
                                     {1.5ex \@plus .2ex}%
                                     {\normalfont\itshape}}
\makeatother

\usepackage{slashed}

\def\+{\!+\!}
\def\-{\!-\!}
\def\={\!=\!}

\begin{document}

\vspace{0.35in}

\begin{center}{\Large{\bf Entanglement entropy and Wilson loop
} } 
\vspace{0.25in}

{\large \bf Bom Soo Kim}\\ ~\\ \vspace{-0.21in}
{\normalsize {Department of Physics, Loyola University Maryland, Baltimore, MD 21210}}\\ 
{\footnotesize  \it  {bkim2@loyola.edu}}\\ 

\end{center}

\vspace{0.11in}

\begin{abstract}
We study both entanglement and the R\'enyi entropies for the 2 dimensional massless Dirac fermions in the presence of topological Wilson loops, which are qualitatively different from those with a chemical potential and a current source. In the language of $\mathbb{Z}_n$ orbifold theories, the Wilson loop is interpreted as an electric operator while the orbifold twist operator as a magnetic operator. Generalized topological transitions for the entropies are driven by both electric and magnetic parameters via the restriction on the operator's conformal weight. By adapting different normalizations for different topological sectors, we achieve two goals: entanglement entropy can be obtained with a smooth limit from the R\'enyi entropy, and the entropies are continuous across the different topological sectors that include general Wilson loops winding sectors. We provide exact results for the entropies in infinite space, which depend only on the topological Wilson loops, independent of the chemical potential and the current source. 
\end{abstract}

%\vspace{0.01in}

{\small\footnotesize
\tableofcontents
}

\newpage 
\section{Introduction}

Entanglement is an important property of quantum theories. Entanglement entropy can be used to measure the way the quantum informations are stored in a quantum state and to classify different quantum states. Its direct computations for the quantum field theories are known to be difficult. Nevertheless there have been steady progresses in 1+1 dimensions \cite{Holzhey:1994we}\cite{Calabrese:2004eu}\cite{Casini:2005rm}\cite{Calabrese:2009qy}\cite{Casini:2009sr} along with various advancements in the holographic entanglement entropies \cite{Ryu:2006bv}\cite{Ryu:2006ef}\cite{Nishioka:2009un}. Exact and analytic results of entanglement entropy for the realistic quantum field systems would be valuable resources for understanding their quantum nature. 

Gauge fields are essential tools for investigating the properties of the quantum fields. Their time and space components are chemical potential and current source (actually source of current), respectively. There have been various scattered results regarding the entropy in the presence of the background gauge fields. For example, at zero temperature, they are independent of a finite chemical potential for free fermions \cite{Ogawa:2011bz}\cite{Herzog:2013py}. For an infinite system with a finite cut, they also have been shown to be independent of chemical potential in \cite{CardySlides:2016}. In recent papers \cite{Kim:2017xoh}\cite{Kim:2017ghc}, we have constructed the general formulas for entanglement and the R\'enyi entropies for the 2 dimensional massless Dirac fermions in the presence of background gauge fields, the chemical potential and current source. These studies show various interesting results and we list some salient features. 
\begin{itemize}
\item Unlike previous results, we have shown that entropies do depend on chemical potential at zero temperature when chemical potential coincides with the energy levels of a quantum system. 
\item In the large radius limit, the entropies do not depend on the gauge fields, supporting a recent claim \cite{CardySlides:2016} and generalizing it to current source and with multi-cuts.
\end{itemize} 

There is an another aspect of the background gauge fields, the topological Wilson loops. The Wilson loops in the context of R\'enyi entropy have been introduced in \cite{Belin:2013uta}. We review this below. The authors of \cite{Belin:2013uta} have computed the R\'enyi entropy on an infinite space using the following definition. 
\begin{align}\label{BelinREE}
S_n (\bar \mu) = \frac{1}{1-n} \log \text{Tr} \left[ \rho_A \frac{e^{\bar \mu Q_A}}{n_A (\bar \mu)} \right]^n \;, 
\end{align}
where $n$ is the number of replica copies, $\bar \mu$ a chemical potential conjugate to $Q_A$, the charge contained in sub-system A and $n_A(\bar \mu) \equiv  \text{Tr} \left[ \rho_A e^{\bar \mu Q_A} \right]$ to ensure that the new density matrix is normalized with the unit trace. It was reported that the R\'enyi entropy computed in \eqref{BelinREE} did not have a smooth entanglement entropy limit for $n\to 1$ \cite{Belin:2013uta}. See also \cite{Goldstein:2017bua} that implements a different normalization and \cite{Matsuura:2016qqu} that has a non-trivial topology in 3 dimensions. 

\begin{figure}[t]
	\begin{center}
		\includegraphics[width=.35\textwidth,angle=-90]{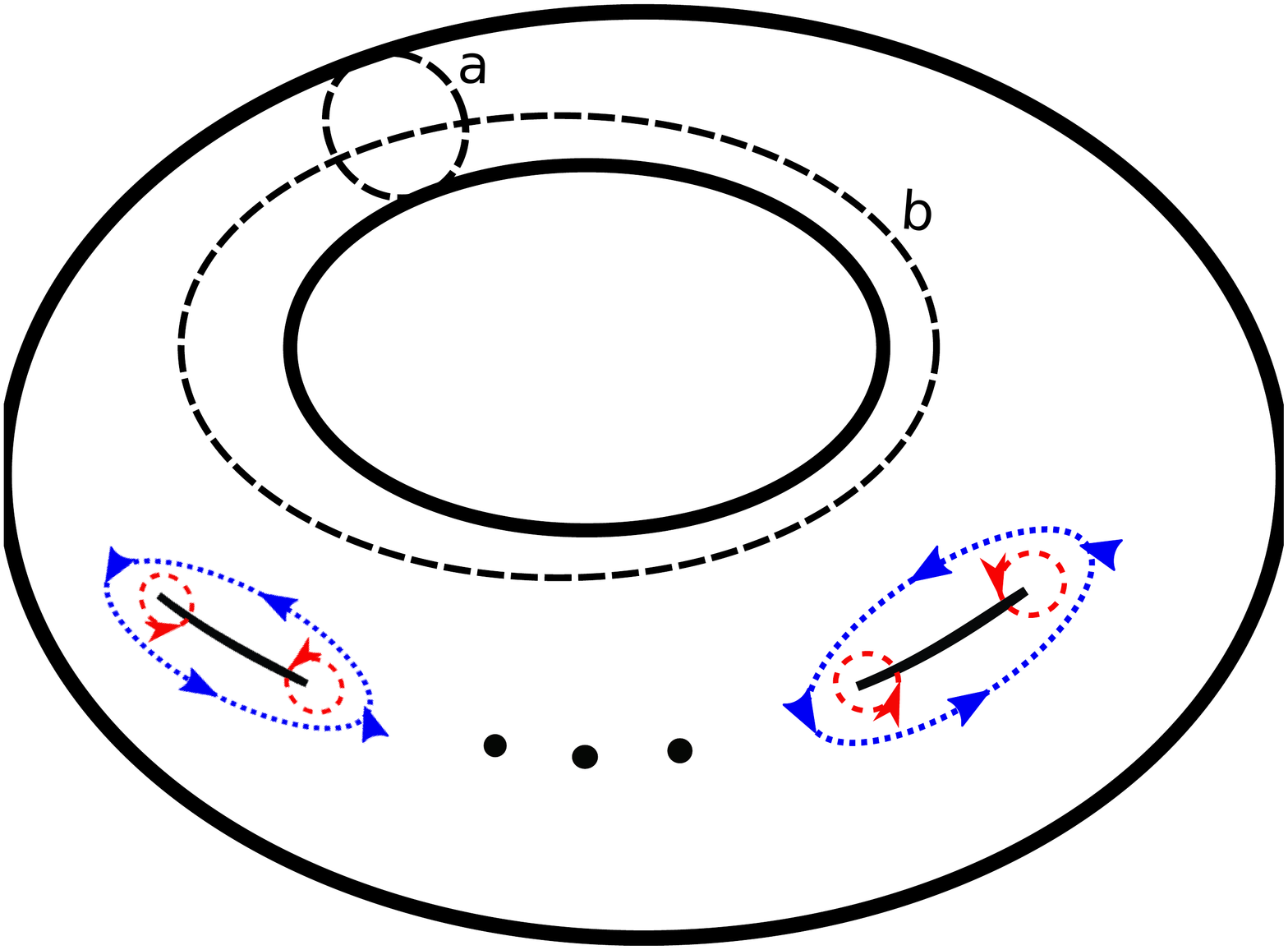} 
		\caption{\footnotesize\small Torus with the two different cycles $a$ and $b$ (represented by the black dashed lines) along with the multiple cuts that have the Wilson loops. On each cut, there are red and blue dotted circles with arrows that represent the electromagnetic twist operators of $\mathbb{Z}_n$ orbifold theories. The red circle represents the magnetic parameter $k= -(n-1)/2, -(n-3)/2, \cdots, (n-1)/2$ of the twist operator, while the blue circle the electric parameter $w$ which is identified as the topological Wilson loops. Chemical potential $\mu$ and currents source $J$ can be understood as parts of the twisted boundary conditions on these two cycles, $a$ and $b$. All these four parameters, $k, w, \mu,$ and $J$, are independent with each other and fully utilized in the context of R\'enyi and entanglement entropies. }
		\label{fig:TorusWithCuts}
	\end{center}
\end{figure}

In this paper, we investigate the effects of the topological Wilson loops to the R\'enyi and entanglement entropies. We generalize previous results in two different ways. First, we expand our previous results \cite{Kim:2017xoh}\cite{Kim:2017ghc} by including the contribution of the Wilson loops to the entropies in a coherent and systematic fashion using the $\mathbb{Z}_n$ orbifold conformal field theories. Twist operators in $\mathbb{Z}_n$ orbifold theories on torus had been well understood. See, {\it e.g.} \cite{DiFrancesco:1997nk}. Previously, we generalized the entropies in the presence of the chemical potential $\mu$ and current source $J$ by realizing that these gauge fields can be understood in terms of twist boundary conditions on the $a$ and $b$ cycles of torus as in the figure \ref{fig:TorusWithCuts}. This identification and generalization of the entropies are reviewed in the \S \ref{sec:PartitionFunction}.

It turns out that the realization of the Wilson loops on torus with cuts are qualitatively distinct from those of the chemical potential and current source. The construction is presented in \S \ref{sec:WilsonLoops} and \S \ref{sec:EEWilsonLoops}. The Wilson loops parameter is nothing but the electric parameter of the electromagnetic vertex operators of the $\mathbb{Z}_n$ orbifold theories \cite{DiFrancesco:1997nk}, while the twist parameter $k/n$, where $k= -(n-1)/2, -(n-3)/2, \cdots, (n-1)/2$ and $n$ is the replica index, represents the operator's magnetic parameter, which was identified in this context in \cite{Azeyanagi:2007bj}. Both the electric and magnetic parameters are parts of the conformal dimension of the electromagnetic twist operators and play key roles in evaluating the entropies. The Wilson loops are represented as the blue dotted lines in the figure \ref{fig:TorusWithCuts}, while the twist parameter of the replica fermion as the red dashed lines at the ends of the cuts. 

We exploit all the possible ingredients of the electromagnetic twist operators of the $\mathbb{Z}_n$ orbifold conformal field theories, and thus the resulting entropy formulas we constructed, \eqref{EEFormula} and \eqref{TopologicalEEFormula1} along with \eqref{TopologicalCorrelatorTwistedOperators}, are comprehensive and exhaustive. 

Second, we generalize the results of \cite{Belin:2013uta} by considering different normalization factors for different Wilson loop sectors and by including all the distinctive Wilson loops winding sectors. This is similar in spirit that the expectation value of an operator is normalized differently for different $\theta$-vacua of QCD \cite{Callen1976}. As explained with details in the introduction \S \ref{sec:WilsonLoops}, the conformal dimensions of the twisted fields depend on the Wilson loops sectors because of the constraint, $-1/2 \leq \alpha_{w,k} = {k}/{n} + {w}/{2\pi} + l_k \leq 1/2$, where $w$ is the Wilson loops parameter. When $w$ increases, the parameter $l_k$ adjusts such that the condition is satisfied. We consider the entire real values for $w$ to cover all the possible winding numbers of the Wilson loops. The changes of the $l_k$ values signify the topological transitions between different sectors, and have practical consequences on the entropies. 

Let us consider the topological transitions of the R\'enyi entropy, that has the form $ S_n = \frac{1}{1- n} ( \log Z[n] - n \log Z[1] )$. We are particularly concerned with the normalization factor $Z[1] $ in the presence of the generalized topological sectors that depend on the parameters $w, n, k $ and $l_k$ through $\alpha_{w,k}$. These topological sectors are completely disconnected, and they do not talk each other. {\it A priori}, there is no clear way to determine the normalization factors across different topological sectors. Thus we employ the following normalization factor   
\begin{align}\label{NormalizationFactorM}
\log Z[1]_w = \lim_{n\to 1}   \left( \log Z[n]_w \right) + \sum_{q=1}^{\infty}  (n-1)^q \log \alpha^w_q
\;, 
\end{align}
where $w$ is the Wilson loops parameter. This form has two advantages. First, the R\'enyi entropy has the smooth entanglement entropy limit, $n\to 1$, that is guaranteed by the first term. The other terms can be used to fine-tune the continuity between different Wilson loops sectors. For example, the linear term $ (n-1) \log \alpha^w_{q=1}$ can be used to have continuous R\'enyi and entanglement entropies. The quadratic term $ (n-1)^2 \log \alpha^w_{q=2}$ can be used to make the first derivative of the entropies be continuous, and so on. This is reminiscent of the various different orders of phase transitions \cite{Charmousis:2010zz}. In \cite{Belin:2013uta}, the authors chose the same normalization for all the topological sectors for the fundamental Wilson loop winding range, $-\pi \leq w < \pi$. Due to that, the R\'enyi entropies do not have the smooth entanglement entropy limits, while the R\'enyi entropies are continuous across different sectors with different $l_k$ within the $w$ range.% 
\footnote{In a previous version of this paper (arXiv:1808.09976v1 on 29 Aug 2018), we employed the first term only. With the choice, the R\'enyi entropies have the smooth entanglement entropies, while the entropies are discontinuous across the different topological sectors. At the moment, all the different possibilities with the different parameters $\alpha^w_q$ seem to be plausible. }      

In this paper, we employ the following normalization factor that seems to be reasonable. 
\begin{align}\label{NormalizationFactorH}
\log Z[1]_w = \lim_{n\to 1}   \left( \log Z[n]_w \right) + (n-1) \log \alpha^w
\;. 
\end{align} 
This is detailed in the main body around the equation \eqref{Normalization} in \S \ref{sec:EESpinIndependent} for the spin structure independent part, around the equation  \eqref{NormalizationSpin} in \S \ref{sec:EESpinDependent}, as well as in the formulations of the R\'enyi entropy in the equation \eqref{TopologicalEEFormula2}. With this choice, our guiding principles are \\ \vspace{-0.2in}

\indent	\qquad I. R\'enyi entropy has a smooth entanglement entropy limit for $n \to 1$. \\
\indent	\qquad II. R\'enyi and entanglement entropies are continuous across different topological sectors. \\  \vspace{-0.2in}

These two conditions, along with the expectation $\log \alpha^w=0 $ for $w=0$, fix the normalization factors completely for all the generalized topological sectors. Different normalization factors will give different results.

We obtain the exact results for infinite space and observe that the entropies depend only the Wilson loops, independent of the chemical potential and current. This is because the spin structure independent parts of the entropies only depend on the conformal dimension of the electromagnetic twist operators. The spin structure dependent parts of the entropies decay as fast as $\mathcal O(\ell_t^2/L^2)$ the infinite space limit $L\to \infty$ with the sub-system size $\ell_t$. 

The paper is organized as follows. We present the general and exhaustive formula for the R\'enyi entropy in the presence of the Wilson loops parameter as well as the chemicla potential and current source in \S \ref{sec:Setup}. We also introduce the conformal dimensions of the electromagnetic operators there. In \S \ref{sec:EESpinIndependent}, we compute the spin structure independent part of the R\'enyi entropy. We discuss a normalization factor that gives us entanglement entropies as smooth limits of the R\'enyi entropies. In \S \ref{sec:EESpinDependent}, we compute the R\'enyi and entanglement entropies for the spin structure dependent part. We consider the high temperature entropies and mutual information in separate sections \S \ref{HighTLimit} and \S \ref{sec:MutualInformation}, respectively. And we conclude in \S \ref{sec:Conclusion}. 

\section{Generalized entanglement entropy with Wilson loops} \label{sec:Setup}

Entanglement entropy can be obtained as a limit from the R\'enyi entropy using the replica trick. The 1+1 dimensional R\'enyi entropy for the massless Dirac fermion can be formulated in a single Riemann surface with $n$-copies of fermions instead of a fermion on $n$-copies of Riemann surfaces \cite{Casini:2005rm}. To include the finite size effect, we consider the torus, and eventually suitable limits are taken for various limits. 

In this language with a single Riemann sheet, the fermions acquire additional twist at both ends of the cut. Thus, the appropriate language for the R\'enyi entropy is the correlation functions of the twist operators of the $\mathbb Z_n$ orbifold theory. Without the gauge fields, the twist operator has only the magnetic parameters $k/n$ for $k$ different fermions. It is straightforward to construct the correlation functions from the partition functions. Thus we start with the partition functions to formulate the R\'enyi entropies in the presence of the Wilson loops as well as the chemical potential and current source in this section. As we see, all these three different gauge fields contribute differently to the entropies.  

\subsection{Partition function with gauge fields} \label{sec:PartitionFunction}

First, we construct the partition function of $1+1$ dimensional Dirac fermions in the presence of current source $ J $ and chemical potential $ \mu $ based on previous studies \cite{DiFrancesco:1997nk}\cite{Hori:2003ic}. In particular, we use the equivalence between the twisted boundary conditions and the background gauge fields to build up the partition function for the fermions. This and the following subsections reveal the clear differences among the Wilson loops, the chemical potential and the current source. 

The action for Dirac fermion $\psi$ in the presence of the gauge field $A_\mu$ is given by  
\begin{align} \label{Action0}
\mathcal S = \frac{1}{2\pi} \int d^2 x ~i   \psi^\dagger \gamma^0  \gamma^\mu \left(\partial_\mu + i A_\mu \right) \psi \;, 
\end{align}
where $\mu=0,1$ is the time or space coordinate, and the corresponding gamma matrices are 
\begin{align}
\gamma^0 = \sigma_1 = \left( \begin{array}{cc}
~0~ & ~1~  \\
1 & 0 \end{array} \right)\;, \qquad  
\gamma^1 = -i \sigma_2 = \left( \begin{array}{cc} 
~0~ & -1~  \\ 
1 & 0 \end{array} \right)\;.
\end{align} 
We also identify the constant background gauge fields as the chemical potential $A_0 = \mu$ and the current source $A_1 = J $. 

The background gauge fields can be identified as twist boundary conditions of the two cycles of the torus, whose modular parameter is $\tau = \tau_1 + i \tau_2$. In the coordinate $\zeta = \frac{1}{2\pi} (s + i t)$, $s$ is the spatial coordinate with a circumference $2\pi L$ (with $L=1$ in this section), and $t$ is the Euclidean time with a periodicity $2\pi \tau_2 = \beta = 1/T$. The space can be identified as $\zeta \equiv \zeta + 1 \equiv \zeta + \tau$.  
For the Dirac field in component form 
\begin{align}
	 \psi  = \left( \begin{array}{c}
	~\psi_-~ \\ 
	\psi_+ \end{array} \right)\;, 
\end{align}
the twisted boundary conditions can be described as 
\begin{align}\label{TwistedeBC} 
\begin{split}
\psi_\pm (t,s) &= e^{-2\pi i a} \psi_\pm (t,s+2\pi) \\
&= e^{-2\pi i b} \psi_\pm (t+2\pi \tau_2, s+2\pi \tau_1)  \;. 
\end{split}
\end{align}

It turns out that there exists an equivalent description between the fermions without the twist boundary conditions in action \eqref{Action0} (with a flat connection $\tilde A_\mu$) and the fermions with the twist boundary condition \eqref{TwistedeBC} and without the gauge fields. These two descriptions can be shown to have the correspondence. 
\begin{align} \label{FlatConnection} 
	\tilde A = \tilde A_\mu dx^\mu 
	= a ds + \frac{b -a\tau_1}{\tau_2} dt \;. 
\end{align}
Thus, one can identify the parameters $a$ and $b$ in terms of the chemical potential and the current source as
\begin{align} 
	a = \tilde J \;, \qquad b 
	= \tau_1 \tilde J + i \tau_2 \tilde \mu \;.
\end{align}
This correspondence greatly simplifies the construction of the partition function for the fermions in the presence of the gauge fields. 

Partition function is nothing but the trace of the Hilbert space that is constructed with the twisted periodic boundary condition for $s \sim s\+ 2\pi$ together with the Euclidean time evolution, $t \to t +2\pi \tau_2$. The latter is represented by the operator $ e^{-2\pi \tau_2 H}$ with the Hamiltonian $H$. In general, the time evolution induces the space translation as well, $s \to s - 2\pi \tau_1$, which is represented by the operator $ e^{-2\pi i \tau_1 P}$, where $P$ is the momentum operator. There is another effect that can come along together for the time evolution. It is a phase rotation due to the presence of the fermion, $ e^{-2\pi i (b - 1/2) F_\psi}$, where $F_\psi = \frac{1}{2\pi} \int ds ( \psi_+^\dagger \psi_+ \- \psi_-^\dagger \psi_-)$ is the fermion number. Putting all together, the partition function with the parameters $a$ and $b$ has the form. 
\begin{align}  
\begin{split}	
	Z_{[a,b]} &= Tr \big[ e^{-2\pi i (b - 1/2) F_\psi} e^{-2\pi i \tau_1 P} e^{-2\pi \tau_2 H} \big] \\
	& = \left|\eta(\tau) \right|^{-2} \big|\vartheta \genfrac[]{0pt}{1}{1/2-a}{b-1/2 } (0,\tau)\big|^2 \;, 
\end{split}	
\end{align}   
where in the second line, we use the Jacobi theta function 
\begin{align} 
	\vartheta \genfrac[]{0pt}{1}{\alpha}{\beta} (z|\tau)  = \sum_{n \in \mathbb Z} q^{\frac{(n + \alpha)^2}{2}}    e^{2\pi i (z+\beta)(n + \alpha)} \;, \qquad 
	q= e^{2\pi i \tau}
\end{align}
and the Dedekind function $\eta(\tau) = q^{\frac{1}{24}} \prod_{n=1}^\infty (1 - q^n)$.

For example, the partition function for the anti-periodic boundary conditions on both circles, $a=1/2, b=1/2$, is given by  
\begin{align}  
	Z_{[\frac{1}{2},\frac{1}{2}]} = \left|\eta(\tau) \right|^{-2} \Big|\vartheta_3 (z|\tau)\Big|^2 
	= \left|\eta(\tau) \right|^{-2} \Big|\sum_{n \in \mathbb Z} q^{\frac{n^2}{2}} \Big|^2 \;. 
\end{align}
In the literature, the following notations are also used. 
\begin{align}
\begin{split}
	&\vartheta_3(z|\tau) = \vartheta \left[\substack{0 \\ 0} \right] (z|\tau) \;, \qquad\quad 
	\vartheta_2(z|\tau) = \vartheta \left[\substack{1/2 \\ 0} \right] (z|\tau) \;,  \\
	&\vartheta_4(z|\tau) = \vartheta \left[\substack{0 \\ 1/2} \right] (z|\tau) \;, \qquad  
	\vartheta_1(z|\tau) = \vartheta \left[\substack{1/2 \\ 1/2} \right] (z|\tau) \;.
\end{split}	
\end{align} 
As mentioned above, $\vartheta_3(z|\tau)$ is related to both the anti-periodic spatial and time circles, while $\vartheta_2(z|\tau)$ is related to the periodic spatial circle and anti-periodic time circle. We focus on these two cases because physical fermions are described by the anti-periodic temporal circle.

In the presence of current source $J$ and chemical potential $\mu$, the partition function can be readily generalized with the equivalence relation we detailed above.  
\begin{align}  
	Z_{[a,b]}^{\mu, J} = Tr \left[e^{2\pi i (\tau_1 J + i \tau_2 \mu + b-\frac{1}{2}) F_A} e^{\-2\pi i \tau_1 P} e^{\-2\pi \tau_2 H} \right] \;.
\end{align}   
If we invoke the equivalent description, this partition function can be understood by the Dirac action that has the periodic boundary conditions in the presence of the background gauge fields $A$ as well as the effective flat gauge connection $\tilde A$.  
\begin{align}  
	\tilde {\mathcal S} = \frac{1}{2\pi} \int d^2 x ~i   \psi^\dagger \gamma^0    \gamma^\mu \left(\partial_\mu + i A_\mu + i \tilde A_\mu\right) \psi \;. 
\end{align}
The background field has various physical effects. In particular, the mode expansions of the fermion depend not only on the twisted boundary condition, but also on the presence of current source $J$. 
\begin{align}  
	\psi_- = \sum_{r \in \mathbb Z +a} \psi_r (t) e^{irs} \qquad  \to  \qquad  \sum_{\tilde r \in \mathbb Z + a + J} \psi_r (t) e^{i \tilde r s} \;.
\end{align}
The partition function in the presence of the chemical potential $\mu$ and current source $J$ along with the boundary conditions can be written as 
\begin{align}  
Z_{[a,b]}^{\mu, J} 
&=\left|\eta(\tau) \right|^{-2} \big|\vartheta \genfrac[]{0pt}{1}{1/2-a- J}{b-1/2 } (\tau_1 J +i\tau_2 \mu|\tau)\big|^2 \;.
\end{align} 
We note that the current source has two distinct contributions: one through the modification of the Hilbert space and another by the periodicity of temporal direction. We also note the corresponding thermal boundary condition is not modified in the presence of the gauge fields! With this we can construct the two point correlation functions in the presence of chemical potential $\mu$ and current source $J$ following \cite{DiFrancesco:1997nk}\cite{Kim:2017ghc}. We present this below, together with the Wilson loops contributions.  

\subsection{Wilson loops contributions} \label{sec:WilsonLoops}

As mentioned above, the R\'enyi entropy can be formulated with $n$-replica copies in the $\mathbb Z_n$ orbifold theory. Here we consider the $n$-replica fermions in the presence of the Wilson loops. Without the Wilson loop, we have the following identifications when the $m$-th fermion $\tilde \psi_m$ ($m=1,2,\cdots, n$) crosses the branch cut connecting two end points, $u$ and $v$, in 2 dimensional Riemann space. 
\begin{align}  
\begin{split}	
	&\tilde \psi_m(e^{2\pi i} (x-u)) = \tilde \psi_{m+1} (x-u) \;, \\ 
	&\tilde \psi_m(e^{2\pi i} (x-v)) = \tilde \psi_{m-1} (x-v) \;,
\end{split}	
\end{align}	  
where $x$ is a complexified coordinate. These can be diagonalized by defining $n$ new fields  
\begin{align}  
	\psi_k = \frac{1}{n} \sum_{m=1}^n e^{2\pi i m k } \tilde \psi_m  \;.
\end{align}	 
For the new diagonalized fields, the boundary conditions develop non-trivial phases. 
\begin{align} 
\begin{split}	
	&\psi_k(e^{2\pi i} (x-u)) = e^{2\pi i k/n} \psi_{k} (x-u) \;, \\
	&\psi_k(e^{2\pi i} (x-v)) = e^{-2\pi i k/n} \psi_{k} (x-v) \;,
\end{split}	
\end{align}	 
where $ k=-(n-1)/2, -(n-3)/2, \cdots, (n-1)/2$. This phase shift $e^{2\pi i k/n}$ signifies that the corresponding twist operator $\sigma_{k/n}$ has a conformal dimension $ k^2/(2 n^2)$. Due to their periodicity, the parameters have the range $-1/2 <  k/n < 1/2$. The full twist operator is the product of these $n$ twist operators, $\sigma_n = \Pi  \sigma_{k/n}$. 

In \cite{Belin:2013uta}, the authors notice that the boundary conditions can be further generalized to include a global phase rotation $\tilde \psi \to e^{i w} \tilde \psi$. This phase can be added to the boundary condition for $\tilde \psi$ as 
\begin{align} 
\begin{split}	
	&\tilde \psi_m(e^{2\pi i} (x-u)) = e^{i w} \tilde \psi_{m+1} (x-u) \;, \\  
	&\tilde \psi_m(e^{2\pi i} (x-v)) = e^{-i w} \tilde \psi_{m-1} (x-v) \;.
\end{split}	
\end{align}	  
It is clear that the diagonal fields has the same effect because this additional phase is added uniformly. 
\begin{align}\label{GeneralTWB} 
\begin{split}	
	&\psi_k(e^{2\pi i} (x-u)) = e^{2\pi i k/n+iw} \psi_{k} (x-u) \;, \\ 
	&\psi_k(e^{2\pi i} (x-v)) = e^{-2\pi i k/n-iw} \psi_{k} (x-v) \;.
\end{split}	
\end{align}	 
We note that these phases have the intrinsic ambiguity under a $2\pi$ shift such that $e^{-2\pi i k/n-iw} = e^{-2\pi i k/n-iw + 2\pi l_k} $. Here we choose the range of periodicity to be $| k/n+w/(2\pi) + l_k| \leq 1/2 $, where $k$ labels different replica fermions. The Wilson loops parameter cover the entire real values to capture all the different winding numbers. As we increase the Wilson loops parameter $w$, the phase becomes bigger than the upper bound. The parameters $l_k$'s can be used to ensure that the combination $k/n+w/(2\pi) + l_k$ stays inside the range. We put a subscript $k$ because $k/n+w/(2\pi)$ is the largest for $ k=(n-1)/2$ and thus the integer $l_k$ depends on the magnetic parameter $k$. 

The non-trivial phases $k/n+w/(2\pi) + l_k$ for the diagonal fields can be handled by the generalized twist operators $\sigma_{w,k}$ with an electric parameter $w$ and a magnetic parameter $k$. Their names come from \cite{DiFrancesco:1997nk} in the language of `electromagnetic' operators for the $\mathbb Z_n$ orbifold theory compactified on a circle, along with the detailed expressions for the correlation functions. The magnetic parameter $\frac{k}{n}$ comes into play when we use the replica trick for $n$ copies of fermions in a single Riemann surface instead of a fermion on $n$ copies of Riemann surfaces \cite{Casini:2005rm}\cite{Casini:2009sr}. The corresponding conformal dimension is given by \cite{DiFrancesco:1997nk}
\begin{align} \label{ConformalDimension} 
	\Delta_{w,k}= \text{conformal dimension} = \frac{1}{2}  \alpha_{w,k}^2= \frac{1}{2} \left( \frac{k}{n} + \frac{w}{2\pi} + l_k \right)^2  \;.
\end{align}	 
The integer constant $l_k$ stems from the fact that the boundary condition \eqref{GeneralTWB} has an intrinsic ambiguity. The $l_k$'s are used to minimize the conformal dimension of the twist operator such that 
\begin{align} \label{AlphaRange}
	-{1}/{2} \leq \alpha_{w,k} < {1}/{2} \;. 
\end{align}
The authors \cite{Belin:2013uta} only considered the range $ -\pi \leq w < \pi$. We choose to cover the entire real number for $w$ because the Wilson loops parameter $w$ already has its own physical meaning as a winding number. Considering only the range $ -\pi \leq w < \pi$ is too restrictive because it only consider the winding number $1$ for the Wilson loops. We generalize the result of \cite{Belin:2013uta} to arbitrary winding numbers. Thus the topological transitions driven by the condition \eqref{AlphaRange} becomes more complicated by two different electric $w$ and magnetic $k$ parameters. 

Now we are ready to write down the expression for the two point correlation functions of the electromagnetic operators $ \sigma_{w,k} (u)$ and $\sigma_{-w,-k} (v)$ with an electric charge $\frac{w}{2\pi} $ and a magnetic charge $\frac{k}{n}$, in the presence of the current source $J$ and the chemical potential $\mu$ as well as the twisted boundary conditions $a, b$ that has been worked out long ago \cite{DiFrancesco:1997nk} 
\begin{align} \label{TopologicalCorrelatorTwistedOperators} 
	\langle \sigma_{w,k} (u) \sigma_{-w,-k} (v) \rangle_{a,b,J,\mu} = \Big| \frac{2\pi \eta (\tau)^3 }{\vartheta [\substack{1/2 \\ 1/2 }](\frac{u-v}{2\pi L}|\tau)} \Big|^{2\alpha_{w,k}^2}~ 
	\Big| \frac{\vartheta [\substack{1/2-a-J \\ b-1/2 }](\frac{u-v}{2\pi L} \alpha_{w,k}+ \tau_1 J + i \tau_2 \mu|\tau)}{\vartheta [\substack{1/2-a-J \\ b-1/2 }](\tau_1 J + i \tau_2 \mu|\tau)} \Big|^2 \;, 
\end{align}
where $\alpha_{w,k} = \frac{k}{n} + \frac{w}{2\pi} + l_k$. Note that we exhaust all the possible ingredients in the context of the $n$-replica fermions, the electric and magnetic parameters $w,k$, chemical potential and current source $\mu,J$, and the twisted boundary parameters $a,b$. This is the most general formula for the correlation functions that are necessary for formulating the R\'enyi entropy. The correlation functions without the Wilson loops in this context can be obtained by setting $w=l_k=0 $ and have been constructed in \cite{Kim:2017xoh}\cite{Kim:2017ghc}. 

\subsection{R\'enyi and entanglement entropy formula} \label{sec:EEWilsonLoops} 

Now we use the correlation functions to formulate the R\'enyi and entanglement entropies. For a sub-system with a cut stretching between two points $u$ and $v$ (with a dimensionless length $(u - v)/2\pi L $), the R\'enyi entropy is given by  
\begin{align} \label{EEFormula}
\begin{split}
S_n &= \frac{1}{1- n} \left[ \log \text{Tr}(\rho_A)^n \right] =\frac{1}{1- n} \Big( \log Z[n] - n \log Z[1] \Big) \\
&\equiv S_n^{w,0} + S_n^{w,\mu,J} \;. 
\end{split}
\end{align}
Where $\rho_A$ is a density matrix. $S_n^{w,0}$ and $ S_n^{w,\mu,J}$ are the spin structure independent and dependent parts, respectively, which are useful for systematic presentations for the rest of the paper. In particular, the results of $S_n^{w,0}$ survives in the infinite space limit. 

Let us connect the definition \eqref{EEFormula} to the advertised correlation functions. 
\begin{align}\label{TopologicalEEFormula1}
	Z[n]= \prod_{k=-{(n-1)}/{2}}^{{(n-1)}/{2}} \langle \sigma_{w,k} (u) \sigma_{-w,-k} (v) \rangle_{a,b,J,\mu} \;. 
\end{align}
The two point functions of the twist operators, $ \sigma_{w,k} (u)$ and $\sigma_{-w,-k} (v)$,  are given in \eqref{TopologicalCorrelatorTwistedOperators}. Note that the parameters $w$ and $k$ are related to the topological transitions discussed in \eqref{ConformalDimension} and \eqref{AlphaRange}. 

Usually, one omits the normalization factor $Z[1]$ in \eqref{EEFormula} because the numerical value turns out to be $1$ and does not contribute to the entropy. This is not the case in the presence of the topological sectors that we consider here. Different topological sectors are disconnected and do not talk to each other. Their corresponding normalization factors do not need to be the same. For example, the expectation values of an operator in different $\theta$-vacua of QCD are normalized differently \cite{Callen1976}. Here we choose the normalization factor to depend on the parameters $w$ and also $k$ through $l_k$.  
\begin{align}\label{TopologicalEEFormula2}
Z[1] = \langle \sigma_{w,0} (u) \sigma_{-w,0} (v) \rangle_{a,b,J,\mu} \;. 
\end{align}
As discussed in the introduction around \eqref{NormalizationFactorM} and detailed below, we encounter the case with a non-trivial normalization factor. There are advantages for choosing different normalizations for different topological sectors. First, it is crucial to have a smooth entanglement entropy limit from the R\'enyi entropy. The paper  \cite{Belin:2013uta} adapted to have the same normalization factors and ended up to have singular limits for the entanglement entropy. Second, one can understand the topological transitions by the R\'enyi entropy by using \eqref{NormalizationFactorM}. The order of topological transitions can be understood with various $\alpha_q^w$ parameters. This specification of $\alpha_q^w$ is one of our main results along with the systematic generalizations of the R\'enyi entropy using the $\mathbb Z_n$ orbifold conformal field theory.  

\section{Entanglement entropy independent of spin structures} \label{sec:EESpinIndependent}

We analyze the spin structure independent entropy $S_n^{w,0}$ in \eqref{EEFormula} in this section. This can be written in a slightly different form. 
\begin{align} \label{EEIndSpinS}
S_n^{w,0} = \frac{1}{1- n} \bigg(  \sum_{k=-{(n-1)}/{2}}^{{(n-1)}/{2}} \alpha_{w,k}^2 - n~ \alpha_{w,0}^2 \bigg) \log \Big|\frac{2\pi \eta (\tau)^3 }{\vartheta [\substack{1/2 \\ 1/2 }](\frac{\ell_t}{2\pi L}|\tau)} \Big|^2   \;.
\end{align}
$S_n^{w,0}$ is independent of chemical potential $\mu$ and current source $J$. We note the second factor $ \alpha_{w,0}^2$ comes from the normalization factor $Z[1]$. When $\alpha_{w,k}$ is independent of $w$, the sum over $k$ can be done in a straightforward manner because $\alpha_{w,k}$ satisfies the stated range \eqref{AlphaRange} and thus $l_k=0$ for all $k$. 

\subsection{Generalized topological transitions}  \label{sec:EESpinIndGenN}

The topological Wilson loops are nothing but the winding number around a cycle. Thus the winding number $\gamma$ sector is associated with the domain $- \gamma \pi \leq w < \gamma \pi$. $\gamma=1$ corresponds to the fundamental domain $ - \pi \leq w < \pi$. As we consider a different winding sector $\gamma=2$, the domain is $- 2\pi \leq w < 2 \pi$. We first focus on the general case with a finite and large $\gamma$ and then specialize some particular sectors. 

In the context of the R\'enyi and entanglement entropies, it is natural to extend this winding sector to the generalized topological sectors. This is driven by the constraint given in \eqref{AlphaRange}. It turns out that the generalized topological transitions are characterized by the change of any $l_k$'s, primarily driven by the magnetic parameter $k$ due to its smaller range, $\Delta (k/n) = 1/n $, compared to the fundamental domain $ \Delta (w/2\pi) < 1$ of the electric parameter $w$. We show this explicitly in this section. 

Now, $\alpha_{w,k}^2$ given in \eqref{ConformalDimension} has an extra dependence on $k$ due to the presence of $l_k$'s, $k = - (n-1)/2, -(n-3)/2, \cdots, (n-1)/2$. Thus the summation, $ \sum_k  \alpha_{w,k}^2 $, cannot be done in a straightforward manner. In particular, the corresponding normalization $ \alpha_{w,0}^2$ for different $l_k$'s used in \cite{Belin:2013uta} produces singular entanglement entropy when one takes $n\to 1$ limit of the R\'enyi entropy. Here we consider the computations more carefully to resolve this issue and to provide a {\it formal} way to compute entanglement entropy from R\'enyi entropy in the presence of the Wilson loop. To derive the general results for the R\'enyi and entanglement entropies, we assume a large and finite number of replica copies. We also provide specific computations for $n=2$ and $n=3$ that provide further consistency. 

The conformal dimensions of the electromagnetic operators \eqref{ConformalDimension} instruct us the condition $ -1/2 \leq \alpha_{w,k} \leq 1/2$ as mentioned above and in \cite{DiFrancesco:1997nk}\cite{Belin:2013uta}. The parameters $l_k$'s are different for different $k$ and also change their values as one increases the value of $w$ that covers the entire real range. The meaning of different Wilson loops sectors with their valid ranges are carefully revisited below. We consider $w>0$, for the entropies for $w<0$ is symmetric to those of $w>0$. 

For $n=2$, there are only two $l_k$'s, $l_{k=1/2}$ and $l_{k=-1/2}$. \\
-- When $0 \leq w <  \pi/2$, all the $l_k$'s vanish, $l_{k=1/2}=l_{k=-1/2}=0$. \\
-- For $\pi/2 \leq w < 3\pi/2 $, $ \alpha_{w,k=1/2} = 1/4 + w/2\pi + l_{k=1/2} $ and $ \alpha_{w,k=-1/2} = -1/4 + w/2\pi + l_{k=-1/2} $. To satisfy the condition $ -1/2 \leq \alpha_{w,k} \leq 1/2$, we need to set $ l_{k=1/2}=-1$ and $ l_{k=-1/2}=0$. \\
-- Similar computations reveal that $ l_{k=1/2}=l_{k=-1/2}=-1$ for $3\pi/2 \leq w < 5\pi/2 $. \\
-- When we increase $w$ further for $5\pi/2 \leq w < 7\pi/2 $, we need $l_{k=1/2}\=-2,  l_{k=-1/2}\=-1. $ \\
-- The next happens with $ l_{k=1/2}\= l_{k=-1/2}\=-2$ for $7\pi/2 \leq w < 9\pi/2 $. The transitions go on with increasing $w$. 

For $n=3$, there are three $l_k$'s, $l_{k=1}$, $l_{k=0}$ and $l_{k=-1}$. When $0 \leq w <  \pi/3$, all the $l_k$'s vanish. For $\pi/3 \leq w < \pi/ $, $ \alpha_{w,k=1} = 1/3 + w/2\pi + l_{k=1} $ and similarly for the other two. To satisfy the condition $ -1/2 \leq \alpha_{w,k} \leq 1/2$, we need to set $ l_{k=1}=-1$ and $ l_{k=0}=l_{k=-1}=0$. Similar computations reveal the following sequence of transitions. $ l_{k=1}=l_{k=0}=-1$ and $ l_{k=-1}=0$ for $\pi \leq w < 5\pi/3 $, $ l_{k=1}=l_{k=0}= l_{k=-1}=-1$ for $5\pi/3 \leq w < 7\pi/3 $. When we increase $w$ further for $7\pi/3 \leq w < 3\pi $, we need $ l_{k=1}=-2$ and $ l_{k=0}=l_{k=-1}=-1$. The transitions go on with increasing $w$. 

These simple cases capture the general pattern for more complicated cases. As $w$ increases staring from $0$, the $l_k$'s develop non-zero values starting for $k=(n-1)/2$ until all the $l_k$'s have the same value, $l_{k=(n-1)/2}= \cdots = l_{k=-(n-1)/2}=-1 $. Further increasing $w$ results in increasing the magnitude of the $l_k$'s starting from  $l_{k=(n-1)/2}=-2$. And the topological transitions continue. We can put these into a general form. 
\begin{itemize}
\item For $0 \leq w <  \pi/n$, all the $l_k$'s vanish. Thus $l_{k=(n-1)/2}\=0,  l_{k=(n-3)/2}\=0, \cdots, l_{k=-(n-1)/2}\=0.$ 

\item For $\pi/n \leq w < 3\pi/n $, one checks the range of $\alpha_{w,k}$'s, especially $ \alpha_{w,k=(n-1)/2} = (n-1)/2n + w/2\pi + l_{k=(n-1)/2} $. Thus, 
$$1/2+ l_{k=(n-1)/2} \leq \alpha_{w,k=(n-1)/2} <  1/2+ 1/n+ l_{k=(n-1)/2}\;.$$ 
To satisfy the condition $ -1/2 \leq \alpha_{w,k} \leq 1/2$, we need to set $ l_{k=(n-1)/2}=-1$. All the other $l_k$'s need to vanish.

\item For $3\pi/n \leq w < 5\pi/n $, one checks the range of $\alpha_{w,k}$'s, especially $ \alpha_{w,k=(n-1)/2} = (n-1)/2n + w/2\pi + l_{k=(n-1)/2} $ and $ \alpha_{w,k=(n-3)/2} = (n-3)/2n + w/2\pi + l_{k=(n-3)/2} $. For $ \alpha_{w,k=(n-3)/2}$, 
$$1/2 + l_{k=(n-3)/2} \leq \alpha_{w,k=(n-3)/2} <  1/2+ 1/n+ l_{k=(n-3)/2}\;.$$
Thus we need to set $ l_{k=(n-3)/2}=-1$.
For $ \alpha_{w,k=(n-1)/2}$, 
$$1/2+ 1/n + l_{k=(n-1)/2} \leq \alpha_{w,k=(n-1)/2} <  1/2+ 2/n+ l_{k=(n-1)/2}\;.$$
Thus we need to impose $ l_{k=(n-1)/2}=-1$. This is valid for $n\geq 2$. Thus we choose 
$$l_{k=(n-1)/2}\=-1, \qquad l_{k=(n-3)/2}\=-1 \;. $$ 
Note that, for $n=2$, further increasing $w$ to the next topological sector $5\pi/2 \leq w < 7\pi/2 $ requires $l_{k=1/2}=-2,  l_{k=-1/2}=-1 $ as explained above. 

\item For $5\pi/n \leq w < 7\pi/n $, one need to check the ranges of $\alpha_{w,k}$ for three different values of $k$, $k=(n-1)/2$, $k=(n-3)/2,$ and $k=(n-5)/2$ for $n\geq 3$. For $k=(n-5)/2$, 
$$ 1/2 + l_{k=(n-5)/2}   \leq \alpha_{w,k=(n-5)/2} < 1/2+1/n + l_{k=(n-5)/2} \;, $$ 
one needs $l_{k=(n-5)/2}=-1 $. 
For $k=(n-3)/2$, 
$$ 1/2+1/n + l_{k=(n-3)/2}   \leq \alpha_{w,k=(n-3)/2} < 1/2+2/n + l_{k=(n-3)/2} \;, $$ 
one needs $l_{k=(n-3)/2}=-1 $. 
For $k=(n-1)/2$, 
$$ 1/2+2/n + l_{k=(n-1)/2}   \leq \alpha_{w,k=(n-1)/2} < 1/2+3/n + l_{k=(n-1)/2} \;, $$ 
one needs $l_{k=(n-1)/2}=-1 $. Thus one needs to impose $ l_{k=(n-1)/2}=l_{k=(n-3)/2}=l_{k=(n-5)/2}=-1$. This continues until we have all the $l_k$'s have $ l_{k=(n-1)/2}=\cdots=l_{k=-(n-1)/2}=-1$, which happens when $w=(2n+1)\pi /n$. 

For $n=3$, the topological sector $ 5\pi/3 \leq w < 7\pi/3 $ requires $ l_{k=1}=l_{k=0}=l_{k=-1}=-1$. Further increase of $w$ to $ 7\pi/3 \leq w < 3\pi $ needs the next level as $ l_{k=1}=-2$ and $l_{k=0}=l_{k=-1}=-1$. This continues. 

\begin{table}[!b]
	\centering
	\begin{tabular}{||c | c | c | c | c | c | c | c|| c | c ||} 
		\hline
		$Q$  & \!$l_{k=(n-1)/2}$\! & \!$l_{k=(n-3)/2}$\! & \!$l_{k=(n-5)/2}$\! & $\cdots \cdots $ & \!$l_{k=(n-5)/2}$\! & \!$l_{k=(n-3)/2}$\! & \!$l_{k=(n-1)/2}$\! & $p$ & $N$ \\ [0.5ex] 
		\hline\hline 
		0 & 0 & 0 & 0 & 0 & 0 & 0 & 0 & 0 & 0 \\ 
		1 & -1 & 0 & 0 & 0 & 0 & 0 & 0 & 1 & 0\\ 
		2 & -1 & -1 & 0 & 0 & 0 & 0 & 0 & 2 & 0\\ 
		3 & -1 & -1 & -1 & 0 & 0 & 0 & 0 & 3 & 0\\ 
		$\vdots$ & $\vdots$ & $\vdots$ & $\vdots$ & $\vdots$ & $\vdots$ & $\vdots$ & $\vdots$ & $\vdots$ & $\vdots$ \\
		$n\!-\!2$ & -1 & -1 & -1 & -1 & -1 & 0 & 0 & $n\!-\!2$ & 0\\
		$n\!-\!1$ & -1 & -1 & -1 & -1 & -1 & -1 & 0 & $n\!-\!1$ & 0\\
		$n$ & -1 & -1 & -1 & -1 & -1 & -1 & -1 & $n$  & 0\\
		$n\!+\!1$ & -2 & -1 & -1 & -1 & -1 & -1 & -1 & $1$ & 1 \\
		$n\!+\!2$ & -2 & -2 & -1 & -1 & -1 & -1 & -1 & $2$ & 1 \\
		$\vdots$ & $\vdots$ & $\vdots$ & $\vdots$ & $\vdots$ & $\vdots$ & $\vdots$ & $\vdots$ & $\vdots$ & $\vdots$\\
		$\vdots$ & $\vdots$ & $\vdots$ & $\vdots$ & $\vdots$ & $\vdots$ & $\vdots$ & $\vdots$ & $\vdots$  & $\vdots$ \\ [1ex] 
		\hline
	\end{tabular}
	\caption{\footnotesize\small General topological transitions for the range, $(2Q-1) \pi/n \leq w < (2Q+1)\pi/n $, with a general parameter $Q$, which is sum over all the $l_k$ values and counts the number of topological transitions. The parametrization $Q=\sum_k |l_k|= Nn+p$ turns out to be convenient below, where $0< p \leq n$ for topologically nontrivial sectors and $N$ counts the number of transitions larger than the replica copies $n$. }
	\label{table:Transitions}
\end{table}

\item \qquad\qquad $\vdots$
\item For $(2n-1) \pi/n \leq w < (2n+1)\pi/n $, one has $l_{k=(n-1)/2}= l_{k=(n-3)/2}=\cdots = l_{k=-(n-1)/2} =-1$. 
\item For $(2n+1) \pi/n \leq w < (2n+3)\pi/n $, one has $l_{k=(n-1)/2}=-2$ with all the rest $-1$. 
\item \qquad\qquad $\vdots$
\item For $(4n-3) \pi/n \leq w < (4n-1)\pi/n $, one has $l_{k=(n-1)/2}=\cdots = l_{k=-(n-3)/2}=-2$, and $ l_{k=-(n-1)/2} =-1$. 
\item For $(4n-1) \pi/n \leq w < (4n+1)\pi/n $, one has $l_{k=(n-1)/2}=\cdots = l_{k=-(n-1)/2} =-2$. 
\item \qquad\qquad $\vdots$ 
\end{itemize}

The table \ref{table:Transitions} summaries the general results of the topological transitions that are indicated by the change of the $l_k$'s.

\subsection{Topological entropies in infinite space: Exact results} 

We compute the R\'enyi and entanglement entropies that are independent of the spin structures and thus survives in the infinite space limit. This section contains the exact results of the entropies that depend on the generalized topological transitions. We construct the entropies for $n=2$ and $n=3$ cases and generalize for arbitrary $n$ in some particular winding sectors along with the general setup. 

\subsubsection{R\'enyi and entanglement entropies $S_n^{w,0}$ for general $n$} \label{sec:SSIndepN}

Due to the complex nature of the entropies, we start with the general $n$ that is assumed to be finite and large. We also consider a winding sector that is also finite and large to avoid specific properties that depend on small $n$ and small winding number. As shown in the previous section, the topological transitions happen when 
\begin{align}
	w = \frac{2Q+1}{n} \pi \;,  
\end{align} 
where $Q$ is an integer. We consider $w \geq 0$ for the entropies are symmetric for $w<0$. This $Q=0, 1, 2, \cdots .$ If one consider a specific winding sector, for example $\gamma=1$ and thus $-\pi \leq w < \pi $, our entropies only defined in that range of $w$ with the end points identified. As we consider our entropies to be continuous, our entropies continues beyond the range for the larger winding sectors. It makes sense to consider the whole range of the parameter space $w$ and specialize for some particular cases later. \\

\noindent {\it $\square$ R\'enyi and entanglement entropies $S_n^{w,0}$ for $w < \pi/n$. } 

Let us evaluate entanglement entropy \eqref{EEIndSpinS} for $w < \pi/n$. We set $l_k=0$ for all $k$ as we discussed. 
\begin{align} \label{SmallwCase}
	\sum_{k=-\frac{n-1}{2}}^{\frac{n-1}{2}} \alpha_{w,k}^2 - n~ \alpha_{w,0}^2 = \sum_{k=-{(n-1)}/{2}}^{{(n-1)}/{2}} \left( \frac{k}{n} + \frac{w}{2\pi} \right)^2 - n \left(  \frac{w}{2\pi} \right)^2 
	  \;.
\end{align}
Note that entanglement and the R\'enyi entropies do not depend on the Wilson loops. In fact, for a small electric parameter $w < \pi/n$, the entropies are the same as those without the Wilson loops. 
\begin{align}\label{RenyiEE0Sector}
S_n^{w<\pi/n,0} = -\frac{n+1}{12 n}  \log \Big|\frac{2\pi \eta (\tau)^3 }{\vartheta [\substack{1/2 \\ 1/2 }](\frac{\ell_t}{2\pi L}|\tau)} \Big|^2 \;.
\end{align}
The corresponding domain becomes $0 \leq w < \pi $, almost covering the fundamental winding domain. We note that this R\'enyi entropy has a smooth entanglement entropy limit when we take $n\to 1$. 

To compare this result to the known results in literature, one can take a large radius limit, $ z= \ell_t/2\pi L \to 0$, along with a zero temperature limit, $q = e^{2\pi i \tau} = e^{-\beta} e^{2\pi i \tau-1} \to 1 $. For that purpose, we use $\vartheta [\substack{1/2 \\ 1/2 }] (z|\tau) = \vartheta_1  (z|\tau) $ and 
\begin{align}
\begin{split}
\vartheta_1  (z|\tau) &= 2 e^{\pi i \tau/4} \sin (\pi z) \prod_{m=1}^\infty (1 - q^m)(1- e^{2\pi i z} q^m)(1- e^{-2\pi i z} q^m)  \sim 2 \sin (\pi z) \sim \ell_t/L  \;, \\
\eta (\tau) &= q^{1/24} \prod_{m=1}^\infty (1 - q^m) \sim 1 \;.
\end{split}
\end{align}
Then, 
\begin{align}\label{RenyiEE0SectorLargeL}
S_n^{w<\pi/n,0} =  \frac{n+1}{6 n} \log |u-v| \qquad \to \qquad S_{n\to 1}^{w<\pi,0} =  \frac{1}{3} \log |u-v| 
\;,
\end{align}
where we use $\ell_t/L \to u-v $ in the large space limit. Our result agrees with \cite{Calabrese:2004eu}\cite{Herzog:2013py}\cite{Belin:2013uta}.   \\

\noindent {\it $\square$ $S_n^{w,0}$ for $\pi/n \leq w < 3\pi/n$. } 

For $\pi/n \leq w < 3\pi/n$, the R\'enyi entropy experiences a transition that requires one of the $l_k$ to be non-zero, $l_{k=(n-1)/2}=-1 $. Thus, 
\begin{align} \label{OneComputation}
\sum_{k=-{(n-1)}/{2}}^{{(n-1)}/{2}} \alpha_{w,k}^2  
&= \sum_{k=-{(n-1)}/{2}}^{{(n-3)}/{2}} \big( \frac{k}{n} + \frac{w}{2\pi} \big)^2 + \big( \frac{n-1}{2n} + \frac{w}{2\pi} - 1 \big)^2 \nonumber \\
&= \sum_{k=-{(n-1)}/{2}}^{{(n-1)}/{2}} \big( \frac{k}{n} + \frac{w}{2\pi} \big)^2 + \big( \frac{n-1}{2n} + \frac{w}{2\pi} - 1 \big)^2 \- \big( \frac{n-1}{2n} + \frac{w}{2\pi} \big)^2 \nonumber \\
&= \frac{n^2-1}{12 n} + n \big(  \frac{w}{2\pi} \big)^2 - \frac{w}{\pi} + \frac{1}{n}
\;.
\end{align}
In \cite{Belin:2013uta}, the normalization was chosen as $\alpha_{w,0}^2= \big(  \frac{w}{2\pi} \big)^2$, which is the same as that of the other sector \eqref{SmallwCase}. Then the R\'enyi entropy \eqref{EEIndSpinS} can be evaluated as 
\begin{align} \label{PreviousResult1}
	S_n^{w,0} = \frac{1}{1- n} ( \frac{n^2-1}{12 n} - \frac{w}{\pi} + \frac{1}{n})  \log \Big|\frac{2\pi \eta (\tau)^3 }{\vartheta [\substack{1/2 \\ 1/2 }](\frac{\ell_t}{2\pi L}|\tau)} \Big|^2 \;. 
\end{align} 
This seems to be fine for the R\'enyi entropy. 

What happens to the entanglement entropy that is associated with \eqref{PreviousResult1}? The limit $n\to 1$ is not well defined, and we end up with a singular entanglement entropy. This difficulty has been discussed in \cite{Belin:2013uta}.

After looking back more carefully, we realize that it is unclear how to define the normalization factor $\alpha_{w,0}^2$ for different topological sectors. {\it In fact, $\alpha_{w,0}^2$ can be chosen differently for different topological sectors.} We choose the following requirements to evaluate the normalization factor.
\begin{itemize}
\item[I.] R\'enyi entropy has a smooth entanglement entropy limit for $n \to 1$. 
\item[II.] R\'enyi and entanglement entropies are continuous across different topological sectors.
\end{itemize}
These two conditions can be met with the following choice. 
\begin{align} \label{Normalization}
	\alpha_{w,0}^2 \equiv \lim_{n \to 1} \Big( \sum_{k=-{(n-1)}/{2}}^{{(n-1)}/{2}} \alpha_{w,k}^2 \Big) + \tilde \alpha_w (n-1)\;.
\end{align}	
The first part ensures that the R\'enyi entropy has a smooth limit for entanglement entropy, while the second part can be used to make the R\'enyi and entanglement entropies be continuous. The spin structure dependent entropies in the following section meet these requirements as well. 

With this in mind, we first evaluate the first part $\alpha_{w,0}^2$ by taking $n\to 1 $ limit in \eqref{OneComputation}. We also demand that the R\'enyi entropy is continuous to \eqref{RenyiEE0Sector} at $w= \pi/n $. This fixes $\tilde \alpha_w =-1/n$. Then, 
\begin{align} \label{TwoComputation}
	\alpha_{w,0}^2 = \big(  \frac{w}{2\pi} \big)^2 - \frac{w}{\pi} + 1 -  \frac{1}{n} (n-1) = \big(  \frac{w}{2\pi} \big)^2 - \frac{w}{\pi} +  \frac{1}{n} 
	\;.
\end{align}
Putting them together, the entropies are 
\begin{align} \label{EEIndSpinSWithp=1}
	S_n^{w,0} = - \bigg( \frac{n+1}{12 n} + \frac{w}{\pi} - \frac{1}{n} \bigg) \log \Big|\frac{2\pi \eta (\tau)^3 }{\vartheta [\substack{1/2 \\ 1/2 }](\frac{\ell_t}{2\pi L}|\tau)} \Big|^2  \qquad \text{for} \qquad \frac{\pi}{n} \leq w < \frac{3\pi}{n}  \;.
\end{align}
Note the entropies do depend on the Wilson loops parameter $w$. This R\'enyi entropy is continuous with \eqref{RenyiEE0Sector} when $w=\pi/n$. Due to the different normalization we take, the results for the R\'enyi and entanglement entropies are different compared to \eqref{PreviousResult1}.  

Entanglement entropy can be obtained by taking $n\to 1$.  
\begin{align} \label{EEIndSpinSWithp=1EE}
S_{n\to 1}^{w,0} =- \bigg( \frac{w}{\pi} - \frac{5}{6} \bigg) \log \Big|\frac{2\pi \eta (\tau)^3 }{\vartheta [\substack{1/2 \\ 1/2 }](\frac{\ell_t}{2\pi L}|\tau)} \Big|^2    \qquad \text{for} \qquad \pi \leq w < 3\pi  \;.
\end{align}
This is a new result. We successfully compute the entanglement entropy in the presence of Wilson loops including the finite size effects that also includes the results of the following section \S \ref{sec:EESpinDependent}. We note that the domain $\pi \leq w < 3\pi $ only covers part of the winding sector $\gamma=2$, while it coincides with the winding sector $\gamma=3$. Actually we already got the final result of the entropies for the $\gamma=3$ winding sector that is the combination of \eqref{RenyiEE0Sector} (with $n\to 1$) and \eqref{EEIndSpinSWithp=1EE} along with the symmetric part of the negative domain, $-3\pi \leq w < 0$. For $\gamma=2$ sector, \eqref{RenyiEE0Sector} covers $0 \leq w < \pi $, while \eqref{EEIndSpinSWithp=1EE} covers $\pi \leq w < 2\pi $, along with their symmetric contributions of negative domains.  \\

\begin{figure}[!th]
	\begin{center}
		\includegraphics[width=.58\textwidth]{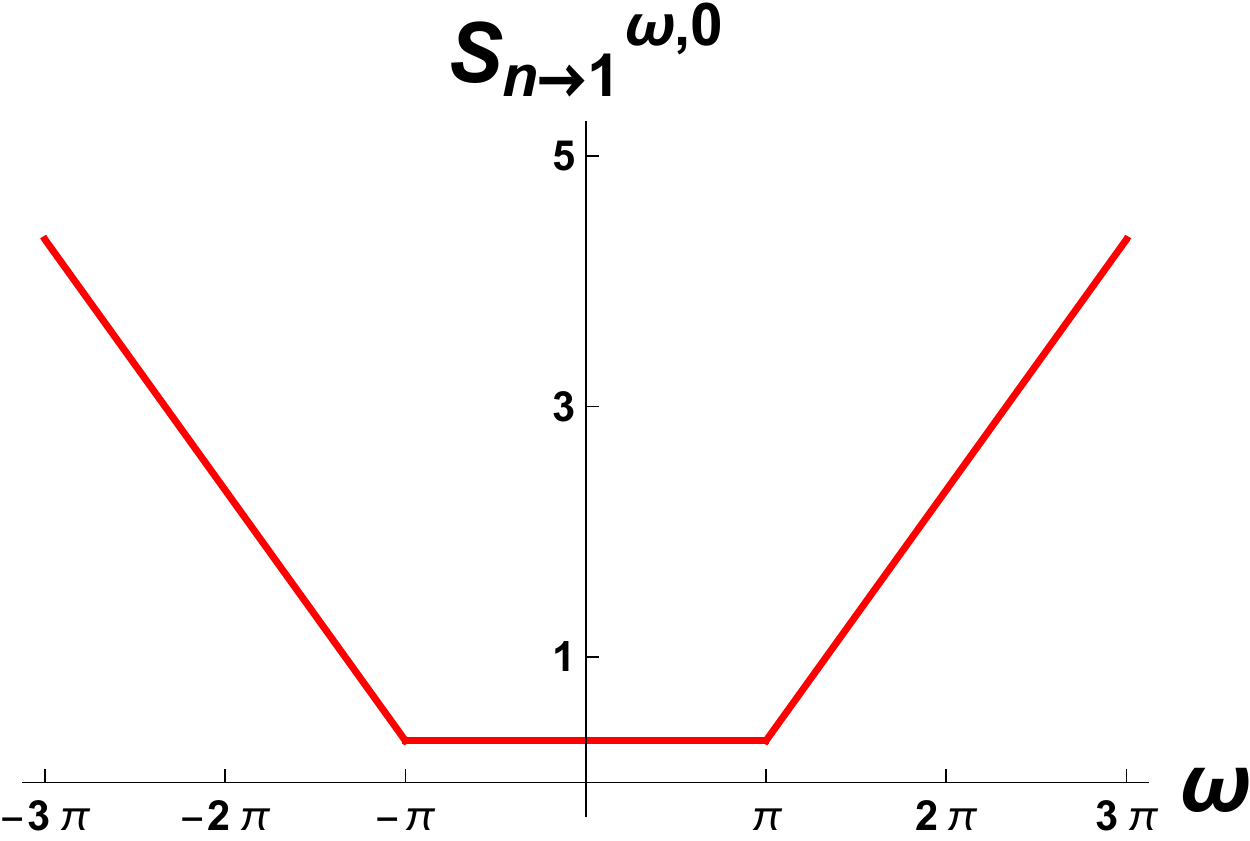} 
		\caption{\footnotesize\small Entanglement entropy for $\gamma=3$ winding sector is combination of \eqref{RenyiEE0SectorLargeL} and \eqref{EEIndSpinSWithp=1EELargeRadius} for $ -3\pi \leq w < 3 \pi$. It is continuous at the transition points $ w=\pm \pi$ as well as $w=\pm 3\pi$, which is the same point due to the periodicity. It turns out that this include the $\gamma=1$ and $\gamma=2$ sectors as well. }
		\label{fig:EESpinIndependentGamma3}
	\end{center}
\end{figure}

The limit for an infinitely large space at zero temperature can be readily evaluated as before. Entanglement entropy becomes 
\begin{align} \label{EEIndSpinSWithp=1EELargeRadius}
S_n^{w,0} = 2 \bigg( \frac{w}{\pi} - \frac{5}{6} \bigg) \log |u-v|   \qquad \text{for} \qquad \pi \leq w < 3\pi  \;.
\end{align}
Where $\ell_t/L = u-v$ is the length of the subsystem. As we increase the Wilson loop parameter $w$, the value of the entanglement entropy increases. This is depicted in figure \ref{fig:EESpinIndependentGamma3} along with \eqref{RenyiEE0SectorLargeL} for the winding sector $\gamma=3$ that covers $-3\pi \leq w < 3\pi$.  \\  

\noindent {\it $\square$ $S_n^{w,0}$ for $3\pi/n \leq w < 5\pi/n$. } 

For $3\pi/n \leq w < 5\pi/n$, the R\'enyi entropy experiences the second topological transition that requires another $l_{k}$'s to be non-zero, $l_{(n-3)/2}=-1 $, in addition to   $l_{(n-1)/2}=-1 $. Thus, 
\begin{align} 
	\sum_{k=-{(n-1)}/{2}}^{{(n-1)}/{2}} \alpha_{w,k}^2  
	&=\sum_{k=-{(n-1)}/{2}}^{{(n-1)}/{2}} \big( \frac{k}{n} + \frac{w}{2\pi} \big)^2 + \big( \frac{n-3}{2n} + \frac{w}{2\pi} - 1 \big)^2 \- \big( \frac{n-3}{2n} + \frac{w}{2\pi} \big)^2 \nonumber  \\
	&\hspace{1.63in} + \big( \frac{n-1}{2n} + \frac{w}{2\pi} - 1 \big)^2 \- \big( \frac{n-1}{2n} + \frac{w}{2\pi} \big)^2  \nonumber \\
	&= \frac{n^2-1}{12 n} + n \big(  \frac{w}{2\pi} \big)^2 - \frac{2w}{\pi} + \frac{4}{n}
	\;.
\end{align}
With the new normalization scheme \eqref{Normalization}, one can work on $\tilde \alpha_w$ so that the R\'enyi entropy is continuous at $w=3\pi/n$. We obtain $ \tilde \alpha_w = -4/n$. 
\begin{align} 
	\alpha_{w,0}^2 = \big(  \frac{w}{2\pi} \big)^2 - \frac{2w}{\pi} + 4 -4 \frac{n-1}{n} = \big(  \frac{w}{2\pi} \big)^2 - \frac{2w}{\pi} + \frac{4}{n}  
	\;.
\end{align}
Using \eqref{EEIndSpinS}, we compute the R\'enyi entropy
\begin{align} \label{EEIndSpinSWithp=2}
	S_n^{w,0} = - \bigg( \frac{n+1}{12 n} + \frac{2w}{\pi} - \frac{4}{n} \bigg) \log \Big|\frac{2\pi \eta (\tau)^3 }{\vartheta [\substack{1/2 \\ 1/2 }](\frac{\ell_t}{2\pi L}|\tau)} \Big|^2 \qquad \text{for} \qquad \frac{3\pi}{n} \leq w < \frac{5\pi}{n}   \;.
\end{align}
Again the entropies are different from those of \cite{Belin:2013uta} due to different normalization. Our results have a smooth limit for entanglement entropy and contain the finite size effects. Furthermore, the entropies are continuous at $ w=3\pi/n$. 

The corresponding entanglement entropy can be readily obtained by taking $n\to 1$ limit of \eqref{EEIndSpinSWithp=2}. Once again, this is the entanglement entropy that contains the winding sector $\gamma=5$ for $-5\pi \leq w < 5\pi$ along with the contributions \eqref{RenyiEE0Sector} and \eqref{EEIndSpinSWithp=1EE} for their corresponding domains. The entanglement entropy for the winding sector $\gamma=4$ is a part of the result of the sector $\gamma=5$.  

For the infinite space limit at zero temperature, we have 
\begin{align} \label{EEIndSpinSWithp=2EELargeLimit}
	S_{n\to 1}^{w,0} = 2 \bigg( \frac{2w}{\pi} - \frac{23}{6} \bigg) \log |u-v|   \qquad \text{for} \qquad 3\pi \leq w < 5\pi   \;.
\end{align}

At this point, one can already get the general picture of evaluating the entropies and the general formula. We still take another example before thinking about the topological transition that involves some of $l_k$'s whose value is beyond $-1$.   \\

\noindent {\it $\square$ $S_n^{w,0}$ for $5\pi/n \leq w < 7\pi/n$.} 

For $5\pi/n \leq w < 7\pi/n$, it is required to have $l_{k=(n-5)/2}=-1$ along with $l_{k=(n-3)/2} = l_{(n-1)/2}=-1 $, which is consistent for $n\geq 3$. Thus, 
\begin{align} 
\sum_{k=-{(n-1)}/{2}}^{{(n-1)}/{2}} \alpha_{w,k}^2  
&=\sum_{k=-{(n-1)}/{2}}^{{(n-1)}/{2}} \big( \frac{k}{n} + \frac{w}{2\pi} \big)^2 + \big( \frac{n-5}{2n} + \frac{w}{2\pi} - 1 \big)^2 \- \big( \frac{n-5}{2n} + \frac{w}{2\pi} \big)^2 \nonumber  \\
&\hspace{1.63in} + \big( \frac{n-3}{2n} + \frac{w}{2\pi} - 1 \big)^2 \- \big( \frac{n-3}{2n} + \frac{w}{2\pi} \big)^2  \nonumber \\
&\hspace{1.63in} + \big( \frac{n-1}{2n} + \frac{w}{2\pi} - 1 \big)^2 \- \big( \frac{n-1}{2n} + \frac{w}{2\pi} \big)^2  \nonumber \\
&= \frac{n^2-1}{12 n} + n \big(  \frac{w}{2\pi} \big)^2 - \frac{3w}{\pi} + \frac{9}{n}
\;.
\end{align}
With the new normalization scheme \eqref{Normalization}, one can work on $\tilde \alpha_w$ so that the R\'enyi entropy is continuous at $w=5\pi/n  $. We obtain $ \tilde \alpha_w = -9/n$. 
\begin{align} 
\alpha_{w,0}^2 = \big(  \frac{w}{2\pi} \big)^2 - \frac{3w}{\pi} + 9 - 9\frac{n-1}{n} = \big(  \frac{w}{2\pi} \big)^2 - \frac{3w}{\pi} + \frac{9}{n}  
\;.
\end{align}

Using \eqref{EEIndSpinS}, we compute the R\'enyi entropy
\begin{align} \label{EEIndSpinSWithp=3}
S_n^{w,0} = - \bigg( \frac{n+1}{12 n} + \frac{3w}{\pi} - \frac{9}{n} \bigg) \log \Big|\frac{2\pi \eta (\tau)^3 }{\vartheta [\substack{1/2 \\ 1/2 }](\frac{\ell_t}{2\pi L}|\tau)} \Big|^2 \qquad \text{for} \qquad \frac{5\pi}{n} \leq w < \frac{7\pi}{n}   \;.
\end{align}
Again our entropies are continuous at $ w=5\pi/n$ and have smooth limits for entanglement entropy. The corresponding entanglement entropy can be evaluated by taking $n \to 1$ limit. For the infinite space limit at zero temperature, we have 
\begin{align} \label{EEIndSpinSWithp=3EELargeLimit}
S_{n\to 1}^{w,0} = 2 \bigg( \frac{3w}{\pi} - \frac{53}{6} \bigg) \log |u-v|   \qquad \text{for} \qquad 5\pi \leq w < 7\pi   \;.
\end{align}

\noindent {\it $\square$ For the general case, $(2Q-1)\pi/n \leq w < (2Q+1)\pi/n$. } 

\begin{figure}[!th]
	\begin{center}
		\includegraphics[width=.58\textwidth]{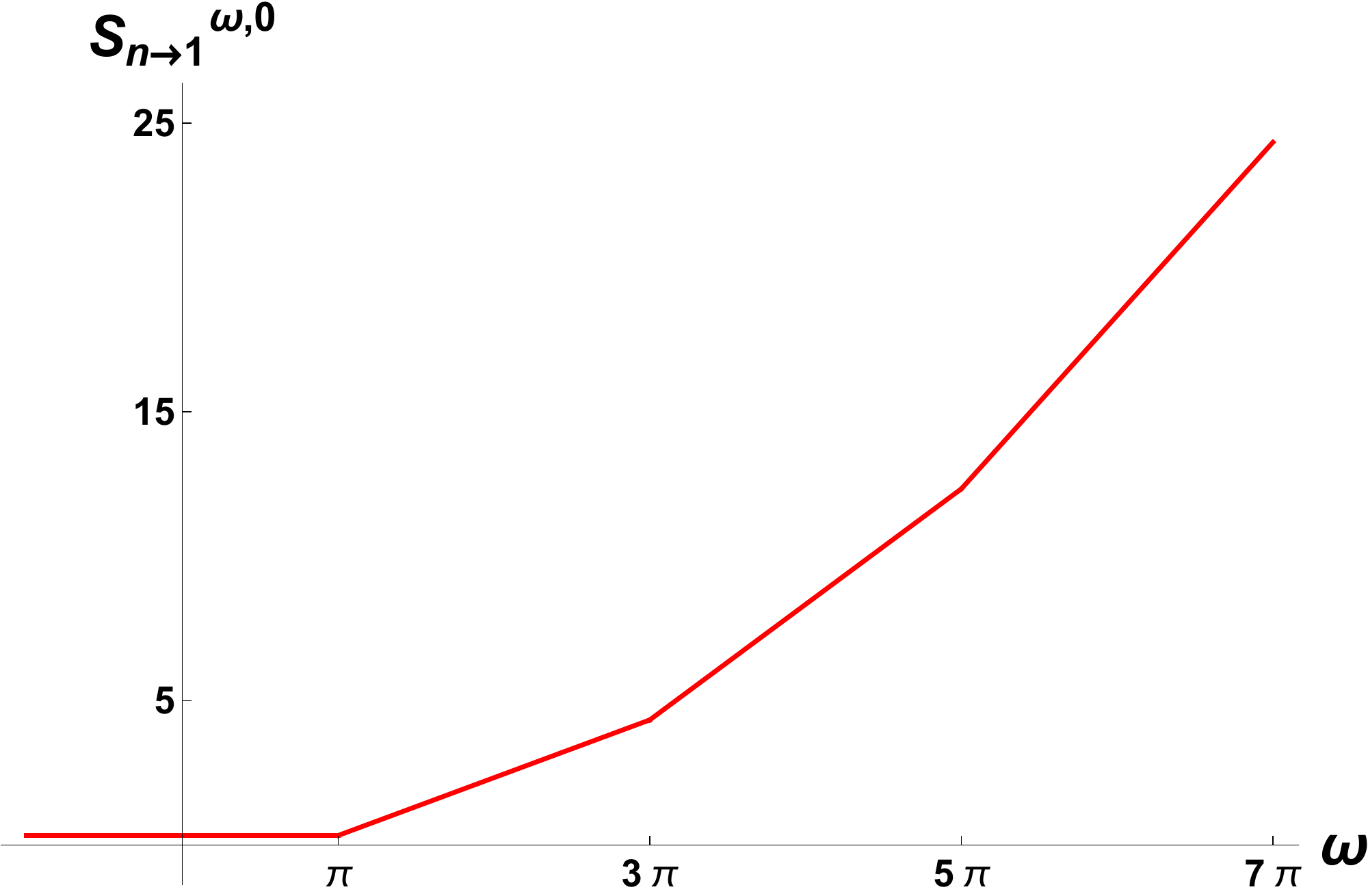} 
		\caption{\footnotesize\small We plot the front factor of \eqref{EEIndSpinSWithPEELargeLimit} for $ 0 \leq w < 7 \pi$. This spin independent part of entanglement entropy is continuous across a different topological sectors and symmetric for $-7\pi \leq w<0$. Its derivatives are discontinuous at $w = (2Q+1) \pi $ for integer $Q$. The plot is linear in $w$ in each topological sector with an interval, $ (2Q-1)\pi \leq w < (2Q+1) \pi$ for a positive integer $Q$. }
		\label{fig:EESpinIndependent}
	\end{center}
\end{figure}

With these examples, we generalize the entropies, whose number of topological transitions is less than or equal to the number of replica copies, $Q=p \leq n$. The number of topological transitions are the number of non-zero $l_k$'s, $Q=p = \sum_{k} |l_k| $ as mentioned in table \ref{table:Transitions}.  
\begin{align} \label{EEIndSpinSWithP}
S_n^{w,0} = - \bigg( \frac{n+1}{12 n} + \frac{p w}{\pi} - \frac{p^2}{n} \bigg) \log \Big|\frac{2\pi \eta (\tau)^3 }{\vartheta [\substack{1/2 \\ 1/2 }](\frac{\ell_t}{2\pi L}|\tau)} \Big|^2 \;,
\end{align}
which is valid for $\frac{(2p-1)\pi}{n} \leq w < \frac{(2p+1) \pi}{n}$. 
For the infinite space limit at zero temperature, we have 
\begin{align} \label{EEIndSpinSWithPEELargeLimit}
S_{n=1}^{w,0} = 2 \bigg( \frac{pw}{\pi} - \frac{6p^2 -1}{6} \bigg) \log |u-v|   \;,
\end{align}
which is valid for $ (2p-1) \pi \leq w < (2p+1)\pi $. Below, we see that these formulas directly generalizes to $Q>n$. Thus, we use $Q$ from now on. This result is depicted in figure \ref{fig:EESpinIndependent}.

Now we argue that these results are true for all the general cases. As we demand the R\'enyi entropies across the topological sectors to be continuous, the entropy is completely fixed once the slope is determined. It turns out that the slope of the entropies in each topological sector is determined by the number of topological transitions. 

This can be easily shown with an example. While we can use the formula specifically true for $n=3$, we can use the general formulas that are also applicable for any $n$. Consider with $n=3$ with a large number of topological transitions $s_1+s_2+s_3 = Q$, where $s_1=|l_{k=1}|$, $s_2=|l_{k=0}|$, $s_3=|l_{k=-1|}$, for $n=3$.  
\begin{align} \label{SpinIndGenComputationP}
\begin{split}
\sum_{k=-{(n-1)}/{2}}^{{(n-1)}/{2}} \alpha_{w,k}^2  
&= \sum_{k=-{(n-1)}/{2}}^{{(n-1)}/{2}} \big( \frac{k}{n} + \frac{w}{2\pi} \big)^2 + \big( \frac{n-5}{2n} + \frac{w}{2\pi} - s_3 \big)^2 - \big( \frac{n-5}{2n} + \frac{w}{2\pi} \big)^2  \\
&\hspace{1.63in}+ \big( \frac{n-3}{2n} + \frac{w}{2\pi} - s_2 \big)^2 - \big( \frac{n-3}{2n} + \frac{w}{2\pi} \big)^2   \\
&\hspace{1.63in} + \big( \frac{n-1}{2n} + \frac{w}{2\pi} - s_1 \big)^2 - \big( \frac{n-1}{2n} + \frac{w}{2\pi} \big)^2   \\
&= \frac{n^2-1}{12 n} + n \big(  \frac{w}{2\pi} \big)^2 -\frac{Qw}{\pi} + (s_1^2+s_2^2+s_3^2 -Q) + \frac{s_1+3s_2+5s_3}{n} 
\;.
\end{split}	
\end{align}
It is not difficult to identify the slope of the resulting entropies as a function of $w$, which is proportional to $Q$. This $Q$ is the number of the topological transitions and sum over the absolute values of all $l_k$'s and also $p$. Thus the slope of the entropies are governed by the number of total topological transitions! One can use this formula to iteratively construct the result \eqref{EEIndSpinSWithP}. In the next subsection, we take an examples to support this result.  

A few comments are in order. First, the results \eqref{EEIndSpinSWithP} and \eqref{EEIndSpinSWithPEELargeLimit} are related to the spin structure independent part of the entropies, $S_n^{w,0}$, and  thus survive in the infinite space limit. As one can check, the entropies have non-trivial dependences on the Wilson loops parameter $w$, while they are independent of the chemical potential and current source. The result \eqref{EEIndSpinSWithPEELargeLimit} is actually exact result in infinite space because the spin dependent part vanishes in the infinite space limit. We also note that the different normalization factors will give different numerical results, yet the fact that the entropies depend on the Wilson loops remains true.  

Second, due to our choice of different normalization factors for different topological sectors, we achieve a smooth entanglement entropy limit as well as the continuous entropies across the different topological sectors. We note that all the possible values are exploited in the sequence of their magnitudes of the $l_k$'s as can be seen in the table \ref{table:Transitions}. This entanglement entropy is plotted in figure \ref{fig:EESpinIndependent}. 

Third, the {\it high temperature limit} can be also computed as in \cite{Kim:2017ghc} using the S-duality relations of the Jacob function. The Jacobi function $ \vartheta [\substack{1/2 \\ 1/2 }] = \vartheta_1$ and the $\eta$ function in \eqref{RenyiEE0Sector} turn into themselves under the S-duality transformation. Thus the final results are the same as \eqref{EEIndSpinSWithP} with the same dependence on $w$ and $p$.

\subsubsection{$n=2$, $n=3$ \& more}

Here we revisit the entropies of $n=2$ and $n=3$ cases in more detail. Here we check that the general results \eqref{EEIndSpinSWithP} and \eqref{EEIndSpinSWithPEELargeLimit} still hold for the transitions with large values of $l_k$'s due to their small numbers. Let us start with $n=3$ case first following the previous subsection \S \ref{sec:EESpinIndGenN} and using \eqref{SpinIndGenComputationP}. \\

\noindent $\square$ When $0 \leq w < \pi/3$, we have $s_1=s_2=s_3=0$. And the entropies are the same as \eqref{RenyiEE0Sector} and \eqref{RenyiEE0SectorLargeL} with $n=3$. Especially the R\'enyi entropy has the form 
\begin{align}\label{RenyiEE0SectorN3}
S_{n=3}^{0\leq w<\pi/3,0} = -\frac{1}{9}  \log \Big|\frac{2\pi \eta (\tau)^3 }{\vartheta _1 ({\ell_t}/{2\pi L}|\tau)} \Big|^2 \;.
\end{align} 

\noindent $\square$ For $\pi/3 \leq w < \pi$, $s_1=1$ and $s_2=s_3=0$. The entropies are straightforward to evaluate. The results are the same as \eqref{EEIndSpinSWithp=1} with $n=3$. The R\'enyi entropy is 
\begin{align} \label{EEIndSpinSWithp=1N3}
S_{n=3}^{\pi/3 \leq w < \pi ,0} = - \big(\frac{w}{\pi} - \frac{2}{9} \big)  \log \Big|\frac{2\pi \eta (\tau)^3 }{\vartheta _1 ({\ell_t}/{2\pi L}|\tau)} \Big|^2  \;.
\end{align}

\noindent $\square$ For $\pi \leq w < 5\pi/3$, $s_1=s_2=1$ and $s_3=0$. The results are \eqref{EEIndSpinSWithp=2} with $n=3$. The R\'enyi entropy is 
\begin{align} \label{EEIndSpinSWithp=2N3}
S_{n=3}^{\pi \leq w < 5\pi/3 ,0} = - \big( \frac{2w}{\pi} - \frac{11}{9} \big)  \log \Big|\frac{2\pi \eta (\tau)^3 }{\vartheta _1 ({\ell_t}/{2\pi L}|\tau)} \Big|^2  \;.
\end{align}

\noindent $\square$ For $5\pi/3 \leq w < 7\pi/3$, $s_1=s_2=s_3=1$. The results are \eqref{EEIndSpinSWithp=3}with $n=3$. The R\'enyi entropy is 
\begin{align} \label{EEIndSpinSWithp=3N3}
\begin{split}
S_{n=3}^{5\pi/3 \leq w < 7\pi/3 ,0} &= - \big( \frac{n+1}{12 n} + \frac{3w}{\pi} - \frac{9}{n} \big)  \log \Big|\frac{2\pi \eta (\tau)^3 }{\vartheta _1 ({\ell_t}/{2\pi L}|\tau)} \Big|^2   \\
&\to - \big( \frac{3w}{\pi} - \frac{26}{9} \big)  \log \Big|\frac{2\pi \eta (\tau)^3 }{\vartheta _1 ({\ell_t}/{2\pi L}|\tau)} \Big|^2  \;.
\end{split}
\end{align}

Until now, we already saw the general procedure to get the result and applied to the specific case $n=3$. From this point, we have more topological transitions than the number of replica copies. Still, the procedure goes smoothly. \\

\noindent $\square$ For $7\pi/3 \leq w < 9\pi/3$, one can check the requirement $|\alpha_k|\leq 1/2$. For $k=1$, $3/2 +l_1 \leq \alpha_1< 5/3 +l_1 $. Thus one needs $l_1=-2 $. Thus, $s_1=2$ and $s_2=s_3=1$. Equation \eqref{SpinIndGenComputationP} with explicit $n$ dependence (even though $n=3$) gives 
\begin{align} 
\sum_{k} \alpha_{w,k}^2 = \frac{n^2-1}{12 n} + n \big(  \frac{w}{2\pi} \big)^2 -\frac{4w}{\pi} + 2 + \frac{10}{n} 
\;.
\end{align}
We choose 
\begin{align} 
\alpha_{w,0}^2 = \big(  \frac{w}{2\pi} \big)^2 -\frac{4w}{\pi} + 12 + \tilde \alpha_w (n-1)  
\;.
\end{align}

The resulting entropy with $n=3$ has the form 
\begin{align} \label{EEIndSpinSN3A}
S_n^{w,0} = - \big( \frac{n+1}{12 n} + \frac{4w}{\pi} - 2 - \frac{10(n+1)}{n} + n \tilde \alpha_w  \big) \log \Big|\frac{2\pi \eta (\tau)^3 }{\vartheta _1 ({\ell_t}/{2\pi L}|\tau)} \Big|^2  \;.
\end{align}
We choose $\tilde \alpha_w$ so that the entropy \eqref{EEIndSpinSN3A} is continuous with \eqref{EEIndSpinSWithp=3N3} at $w= 7\pi/n \to 7\pi/3$. This fixes 
\begin{align}
	n\tilde \alpha_w = -12  + \frac{6}{n} \;.
\end{align}
Thus we get 
\begin{align}
\begin{split} 
S_{n=3}^{7\pi/3 \leq w < 9\pi/3 ,0} &= - \big( \frac{n+1}{12 n} + \frac{4w}{\pi} - \frac{16}{n} \big)  \log \Big|\frac{2\pi \eta (\tau)^3 }{\vartheta _1 ({\ell_t}/{2\pi L}|\tau)} \Big|^2 \\
&\to  - \big( \frac{4w}{\pi} - \frac{47}{9} \big)  \log \Big|\frac{2\pi \eta (\tau)^3 }{\vartheta _1 ({\ell_t}/{2\pi L}|\tau)} \Big|^2 \;.
\end{split}
\end{align}
This example shows that the topological transitions with $|l_k|>2$ is actually smooth, and our general formulas \eqref{EEIndSpinSWithP} and \eqref{EEIndSpinSWithPEELargeLimit} hold. 

This procedure can continue to build the rest of the R\'enyi entropy. For example, $s_1=s_2=s_3=2$ or $s_1 =5, s_2= s_3=4$. Note that we do not have the combination $s_1=2, s_2=s_3=3$ because there are constraints $s_1 \geq s_2 \geq s_3 $ and they differ by $1$ at the most according to table \ref{table:Transitions} as $ s_1-s_2 \leq 1$ and $ s_2-s_3\leq 1$. 

Thus we again arrive the same general result.  
\begin{align} \label{RenyiIndSpinSWithQ}
S_n^{\frac{(2Q-1)\pi}{n} \leq w < \frac{(2Q+1) \pi}{n},0} = - \big( \frac{n+1}{12 n} + Q\frac{w}{\pi} - Q^2 \frac{1}{n} \big) \log \Big|\frac{2\pi \eta (\tau)^3 }{\vartheta _1 ({\ell_t}/{2\pi L}|\tau)} \Big|^2 \;.
\end{align}
Where $Q = \sum_k |l_k| $ measures the number of the topological transitions even when $|l_k|>1$. This formula applies for appropriate $n$, which is true for $n=3$ in this section. As we keep the $n$ dependence with full generality, we also take the limit $n\to 1$ as well to get the entanglement entropy limit. 
\begin{align} \label{EEIndSpinSWithQ}
S_{n\to 1}^{(2Q-1)\pi \leq w < (2Q+1) \pi,0} = - \big(  Q\frac{w}{\pi} - Q^2 + \frac{1}{6} \big) \log \Big|\frac{2\pi \eta (\tau)^3 }{\vartheta _1 ({\ell_t}/{2\pi L}|\tau)} \Big|^2 \;.
\end{align}
It is interesting to see how $n\to 1$ limit emerges as we lower $n$. Thus we consider the details for $n=2$.  \\

For $n=2$, there are only $s_1=|l_{1/2}|$ and $s_2=|l_{-1/2}|$. From \eqref{SpinIndGenComputationP}, one can set $s_3=0$ to get 
\begin{align} \label{SpinIndGenComputationPN2}
\begin{split}
\sum_{k=-{(n-1)}/{2}}^{{(n-1)}/{2}} \alpha_{w,k}^2  
&= \sum_{k=-{(n-1)}/{2}}^{{(n-1)}/{2}} \big( \frac{k}{n} + \frac{w}{2\pi} \big)^2 + \big( \frac{n-3}{2n} + \frac{w}{2\pi} - s_2 \big)^2 - \big( \frac{n-3}{2n} + \frac{w}{2\pi} \big)^2   \\
&\hspace{1.63in} + \big( \frac{n-1}{2n} + \frac{w}{2\pi} - s_1 \big)^2 - \big( \frac{n-1}{2n} + \frac{w}{2\pi} \big)^2   \\
&= \frac{n^2-1}{12 n} + n \big(  \frac{w}{2\pi} \big)^2 -\frac{Qw}{\pi} + (s_1^2+s_2^2  -Q) + \frac{s_1+3s_2}{n} 
\;,
\end{split}	
\end{align}
where $Q=s_1+s_2$. As already mentioned, we build our R\'enyi entropies with the principle of continuity across the topological sectors. Only the slope matters as we see here explicitly. We also set the starting point at $w=0$ with $Q=0$ that is equivalent to the entropies without Wilson loops. \\

\noindent $\square$ When $0 \leq w < \pi/2$, we have $s_1=s_2=0$. And the entropies are the same as \eqref{RenyiEE0Sector} and \eqref{RenyiEE0SectorLargeL} with $n=2$. This result can be iteratively constructed without referring to these general results. 
\begin{align}\label{RenyiEE0SectorN2}
S_{n=2}^{0\leq w<\pi/2,0} = -\frac{1}{8}  \log \Big|\frac{2\pi \eta (\tau)^3 }{\vartheta _1 ({\ell_t}/{2\pi L}|\tau)} \Big|^2 \;.
\end{align} 

\noindent $\square$ For $\pi/2 \leq w < 3\pi/2$, $s_1=1$ and $s_2=0$. The results are the same as \eqref{EEIndSpinSWithp=1} with $n=2$. Thus,
\begin{align} \label{EEIndSpinSWithp=1N2}
S_{n=2}^{\pi/2 \leq w < 3\pi/2 ,0} = - \big(\frac{w}{\pi} - \frac{3}{8} \big)  \log \Big|\frac{2\pi \eta (\tau)^3 }{\vartheta _1 ({\ell_t}/{2\pi L}|\tau)} \Big|^2  \;.
\end{align}

\noindent $\square$ For $3\pi/2 \leq w < 5\pi/2$, $s_1=s_2=1$. The results are \eqref{EEIndSpinSWithp=2} with $n=2$. The R\'enyi entropy is 
\begin{align} \label{EEIndSpinSWithp=2N2}
S_{n=2}^{3\pi/2 \leq w < 5\pi/2 ,0} = - \big( \frac{2w}{\pi} - \frac{15}{8} \big)  \log \Big|\frac{2\pi \eta (\tau)^3 }{\vartheta _1 ({\ell_t}/{2\pi L}|\tau)} \Big|^2  \;.
\end{align}

\noindent $\square$ For $5\pi/2 \leq w < 7\pi/2$, $s_1=2$ and $ s_2=1$. The results are \eqref{EEIndSpinSWithp=3} with $n=2$. The R\'enyi entropy is 
\begin{align} \label{EEIndSpinSWithp=3N2}
\begin{split}
S_{n=2}^{5\pi/2 \leq w < 7\pi/2, 0} &= - \big( \frac{n+1}{12 n} + \frac{3w}{\pi} - \frac{9}{n} \big)  \log \Big|\frac{2\pi \eta (\tau)^3 }{\vartheta _1 ({\ell_t}/{2\pi L}|\tau)} \Big|^2   \\
&\to - \big( \frac{3w}{\pi} - \frac{35}{8} \big)  \log \Big|\frac{2\pi \eta (\tau)^3 }{\vartheta _1 ({\ell_t}/{2\pi L}|\tau)} \Big|^2  \;.
\end{split}
\end{align}

The following topological transition happens for $7\pi/2 \leq w < 9\pi/2$, $s_1=s_2=2$ and thus with $Q=4$. One can easily figure out the R\'enyi and entanglement entropies using \eqref{RenyiIndSpinSWithQ} and \eqref{EEIndSpinSWithQ} for any number of transitions $Q$.   

\begin{figure}[!h]
	\begin{center}
		\includegraphics[width=.56\textwidth]{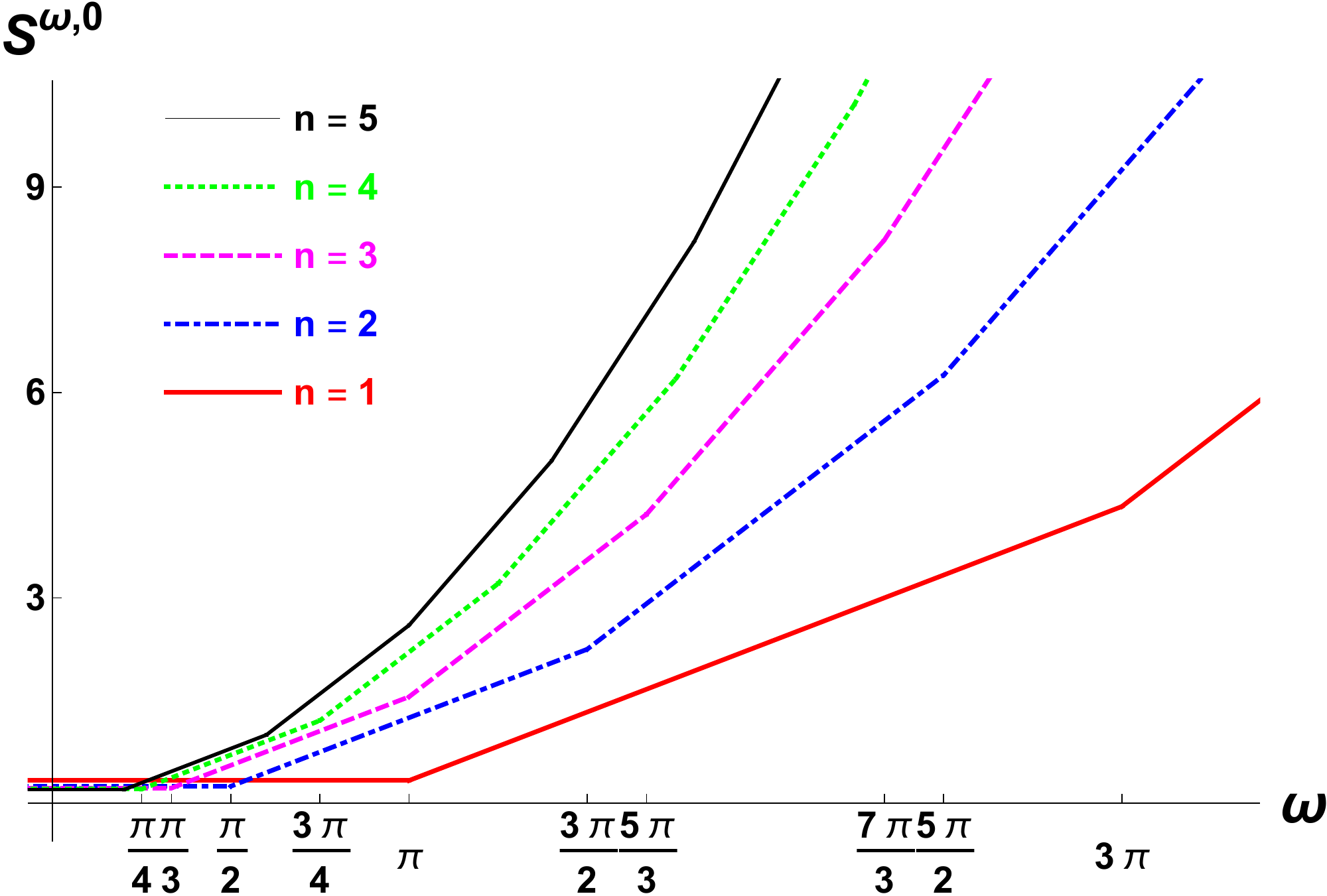} 
		\caption{\footnotesize\small Plots for the entropies $n=5,4,3,2,1$. We see the transition points for $n=1$ at $w=\pi, 3\pi, \cdots$ in the bottom red curve, for $n=2$ at $w=\pi/2, 3\pi/2, 5\pi/2, \cdots $ in the blue dot-dashed curve, for $n=3$ at $w=\pi/3, \pi, 5\pi/3, 7\pi/3, \cdots$, and so on. }
		\label{fig:EESpinIndepPEELimit}
	\end{center}
\end{figure}

After figuring out the details of $n=2$ and $n=3$, we can easily generalize the entropies for other values of replicas. Instead of writing them down, we depict them in the figure \ref{fig:EESpinIndepPEELimit} for $n=1,2,3,4,5$. As mentioned, the slope of the entropies depends on the number of transitions $Q$ and is independent of $n$. This is clearly visible in the figure. The slopes of the entropies are $0$ for the case with $Q=0$ that happens in the legs containing $w=0$. After the first transition happening $w=\pi/n $ for given $n$, all the entropies have the same slope.  

From the figure \ref{fig:EESpinIndepPEELimit}, we can pictorially see the way to take the entanglement entropy limit, $n\to 1$. As we take the limit, each topological sector enlarges. The $Q=0$ sector coincides with the $\gamma=1$ winding sector. $Q=1$ sector encompass $-3\pi \leq w < 3\pi$, which is $\gamma=3$ winding sector. $Q=2$ sector includes $-5\pi \leq w < 5\pi$, which is $\gamma=5$ winding sector. And so on. 

If one considers only $\gamma=2$ winding sector from the beginning, only the region $ -2\pi \leq w < 2\pi$ is visible. This is perfectly fine because the entropies are continuous at $ w=\pm 2\pi$. For $n=3$, there are three topological transitions in this region, while there is only one for $n=1$.   

Finally, we mention that all these contributions survive in the infinite space limit! In that limit, the entropies \eqref{RenyiIndSpinSWithQ} and \eqref{EEIndSpinSWithQ} become exact because the spin structure dependent entropies vanish.

\newpage 
\section{Entanglement entropy depending on spin structures} \label{sec:EESpinDependent}

In this section, we compute spin structure dependent R\'enyi and entanglement entropies $S_n^{w,\mu,J}$ given in \eqref{EEFormula}. We emphasize the role of the Wilson loops parameter $w$, which has been identified as an electric parameter of the electromagnetic twist operators. 
\begin{align} \label{TopologicalEEFormulaJmu0}
S_n^{w,\mu,J}
&= \frac{1}{1- n} \Big[ \log \tilde Z_{s}[n]  -  n \log \tilde Z_{s}[1] \Big] \;, 
\end{align} 
where $\tilde Z_{s}[n]$, a part of $Z[n]$ that depends on the spin structures, can be written in terms of the Jacobi theta functions as 
\begin{align} \label{TopologicalEEFormulaJmu1}
\tilde Z_{s}[n] &= \prod_{k=-{(n-1)}/{2}}^{{(n-1)}/{2}} \Big| \frac{\vartheta [\substack{1/2-a-J \\ b-1/2 }](\frac{\ell_t}{2\pi L} \alpha_{w,k} + \tau_1 J + i \tau_2 \mu|\tau)}{\vartheta [\substack{1/2-a-J \\ b-1/2 }](\tau_1 J + i \tau_2 \mu|\tau)} \Big|^2  \;.
\end{align} 

As mentioned in the previous section, determining normalization factors is not straightforward. We follow the same requirements as in the previous section. 
\begin{itemize}
	\item[I.] R\'enyi entropy has a smooth entanglement entropy limit for $n \to 1$. 
	\item[II.] R\'enyi and entanglement entropies are continuous across different topological sectors.
\end{itemize}
These two conditions can be met with the following choice. 
\begin{align} \label{NormalizationSpin}
\log \tilde Z_{s}[1] = \lim_{n\to 1}   \left( \log \tilde Z_{s}[n] \right) + (n-1) \log \tilde \alpha_{s}  \;.
\end{align} 
The first part on the right hand side guarantees the requirement I, while the second part guarantees the continuity of the R\'enyi and entanglement entropies, the requirement II. 

The Wilson loops $w$ dependence of the R\'enyi and entanglement entropies on a torus has not been studied in the literature. The Jacobi theta functions in \eqref{TopologicalEEFormulaJmu1} have the $w, n, k, \mu$ and $J$ dependences in the argument. Thus the entropies depend on the Wilson loops, chemical potential, and current source. We also note that these spin structure dependent entropies vanish in infinite limit. 

Obtaining closed form analytically by summing over $k$ with full generality is not feasible. The case with $w=0$ has been throughly studied in various limits, such as the zero temperature limit and the large radius limit, in \cite{Kim:2017ghc}. There we showed novel and interesting results. For more details, refer to \cite{Kim:2017ghc}. 

\subsection{Entanglement entropy $S_n^{w,\mu,J}$ with $\mu =J =0$. } 

To simplify the computations and see the role of the Wilson loops on the entropies clearly, we choose $ \mu\=J\=0$. We consider the general case with non-zero $\mu$ and $J$ in the next section. We focus on the anti-periodic fermions by setting $a\=b\={1}/{2}$. Periodic fermions on the spatial circle can be understood in a similar manner. The entropies can be computed by putting $ \mu\=J\=0$ in \eqref{TopologicalEEFormulaJmu1}. 
\begin{align}  \label{TopologicalJMUW22}
	\log \tilde Z_{s}[n] = \sum_{k=-{(n-1)}/{2}}^{{(n-1)}/{2}} \log \Big| \frac{\vartheta_3 (\frac{\ell_t}{2\pi L} [\frac{k}{n} + \frac{w}{2\pi} + l_k]|\tau)}{\vartheta_3(0|\tau)} \Big|^2 
	\;.
\end{align} 
Before the actual computations, we mention that the topological transition patterns follow table \ref{table:Transitions}. We consider a large and finite replica copies, $n$, for a large and finite Wilson loop number sector, $\gamma$, so that we can understand the full picture. We illustrate some specific examples with a figure. Detailed computations for entropies with similar forms have been presented in \cite{Kim:2017ghc}. Thus we quote the results for various limits by emphasizing the role of $l_k$'s, except the computations related to the Wilson loops parameter.   

\subsubsection{Low temperature limit, Anti-periodic fermion} \label{sec:FiniteEEMuJZero}

\noindent {\it $\square$ R\'enyi and entanglement entropies $S_n^{w,\mu=J=0}$ for $0 \leq w < \pi/n$. } \\
First, we consider $w< \pi/n$ in the low temperature limit $\beta=2\pi \tau_2 \to \infty$. For the theta functions in \eqref{TopologicalJMUW22}, we use $|q|=|e^{2\pi i \tau}| \ll 1$  to expand $\vartheta_3 (z|\tau) = \prod_{m=1}^\infty (1 - q^m)(1 + y q^{m-1/2})(1 + y^{-1} q^{m-1/2})$ for general $y = e^{2\pi i z} $. The sums over $k$ and $m$ can be done independently. We find
\begin{align}  \label{LowTComputationSmallw}
	\log \tilde Z_{s}[n] 
	&= 4\sum_{k=-{(n-1)}/{2}}^{{(n-1)}/{2}} \sum_{l,m=1}^{\infty}  \frac{(-1)^{l-1}}{l} 
	\frac{\cos (\alpha l (m-1/2)) }{e^{l\beta (m-1/2 )}} \Big [ \cos \Big( \frac{\ell_t l}{L} \big[\frac{k}{n} + \frac{w}{2\pi} + l_k \big] \Big) - 1 \Big]  \nonumber  \\
	&=4 \sum_{l=1}^{\infty} \frac{(-1)^{l-1}}{l} 
	\frac{\sinh (\frac{l\beta }{2}) \cos (\frac{l \alpha }{2})}{\cosh ( l\beta ) - \cos (l \alpha)} \Big[ \cos \left( \frac{\ell_t l}{2\pi L} w\right) \frac{\sin (\frac{\ell_t l}{2L} )}{\sin (\frac{\ell_t l}{2L n} )}  -n \Big]
	\;. 
\end{align} 
The parameter $\alpha=2\pi \tau_1$ is different from $\alpha_{w,k}$. 

For the normalization factor \eqref{NormalizationSpin}, the first part on the right hand side is straightforward to give
\begin{align}  
\begin{split}
	\log \tilde Z_{s}[1] &=4 \sum_{l=1}^{\infty} \frac{(-1)^{l-1}}{l} 
	\frac{\sinh (\frac{l\beta }{2}) \cos (\frac{l \alpha }{2})}{\cosh ( l\beta ) - \cos (l \alpha)} \Big[ \cos \big( \frac{\ell_t l}{2\pi L} w \big)  -1 \Big] \\
	&+ (n-1) \log \tilde \alpha_{s}
	\;. 
\end{split}	
\end{align} 
We choose $\tilde \alpha_{s}=1$ and thus $ \log \tilde \alpha_{s}=0 $ because the phase $l_k$'s are trivial for $w< \pi/n$. R\'enyi entropy becomes
\begin{align} \label{LowTRenyiSmallW}
	S_n^{w, \mu=J=0}
	&= \sum_{l=1}^{\infty} \frac{(-1)^{l-1}}{l} 
	\frac{\sinh (\frac{l\beta }{2}) \cos (\frac{l \alpha }{2})}{\cosh ( l\beta ) - \cos (l \alpha)} \cos \Big( \frac{\ell_t l}{2\pi L} w \Big)
	\frac{4}{1\- n} \Big[  \frac{\sin (\frac{\ell_t l}{2L} )}{\sin (\frac{\ell_t l}{2L n} )}  -n \Big] \;,
\end{align}
which is valid for $ 0\leq w< \pi/n$. There is smooth $n\to 1$ limit that gives entanglement entropy. 
\begin{align} \label{LowTEE00Small}
	S_{n\to 1}^{0\leq w< \pi/n, \mu=J=0}
	&= 4 \sum_{l=1}^{\infty} \frac{(-1)^{l-1}}{l} 
	\frac{\sinh (\frac{l\beta }{2}) \cos (\frac{l \alpha }{2})}{\cosh ( l\beta ) - \cos (l \alpha)} \cos \Big( \frac{\ell_t l}{2\pi L} w \Big)
	\Big[ 1 - \frac{\ell_t l}{2L} \cot (\frac{\ell_t l}{2L} ) \Big] 
	\;.
\end{align} 
Note that the R\'enyi and entanglement entropies depend on the Wilson loop parameter as oscillating functions $ \cos \left( \frac{\ell_t l}{2\pi L} w\right)$ for small $w$. This is a finite size effect. For infinite space $L\to \infty$, $ \cos \left( \frac{\ell_t l}{2\pi L} w\right) \to 1$. Moreover, the terms in square bracket behave as $1 - \frac{\ell_t l}{2L} \cot (\frac{\ell_t l}{2L} ) \sim \frac{1}{3}(\frac{\ell_t l}{2L} )^2  $, which vanishes in infinite space. Thus, the $w$ dependence of the entropies drops out. The rest of the results are consistent with the previous results \cite{Kim:2017ghc}.  \\

\noindent {\it $\square$ R\'enyi and entanglement entropies $S_n^{w,\mu=J=0}$ for general case, $(2Q-1)\pi/n \leq w < (2Q+1)\pi/n$. } \\
One can compute the entropies for each topological sector iteratively starting $\pi/n \leq w < 3\pi/n$, and build up for the general case $(2Q-1)\pi/n \leq w < (2Q+1)\pi/n$ with an integer $Q$. The computations are cumbersome, yet straightforward. 

We start with the number of the topological transitions are smaller than the replica copies, $Q=p \leq n$, so that $l_k=-1, 0$ for all $k$.   
\begin{align}  \label{LowTComputationGeneralpW}
\log \tilde Z_{s}[n] 
&= 4 \sum_{l,m=1}^{\infty}  \frac{(-1)^{l-1}}{l} 
\frac{\cos (\alpha l (m-1/2)) }{e^{l\beta (m-1/2 )}} \bigg[ \sum_{k=-\frac{n-1}{2}}^{\frac{n-1}{2}} \cos \Big( \frac{\ell_t l}{L} \big[\frac{k}{n} + \frac{w}{2\pi} \big] \Big) - n \bigg.    \\
& + \sum_{r=1}^p \bigg\{ \cos \Big( \frac{\ell_t l}{L} \big[\frac{n\-(2r-1)}{2n} + \frac{w}{2\pi} - 1 \big] \Big)  -  \cos \Big( \frac{\ell_t l}{L} \big[\frac{n-(2r-1)}{2n} + \frac{w}{2\pi}  \big] \Big) \bigg\}  \bigg.  \bigg] \;. \nonumber
\end{align} 
The first line is the same as \eqref{LowTComputationSmallw} and so are the entropies. The second line contains all the new terms due to the topological transitions. For example, $p=1$ case has $\cos \left( \frac{\ell_t l}{L} \big[\frac{n -1}{2n} + \frac{w}{2\pi} - 1 \big] \right) - \cos \left( \frac{\ell_t l}{L} \big[\frac{n-1}{2n} + \frac{w}{2\pi} \big] \right) $, which is relevant for $\pi/n \leq w < 3\pi/n$. The sum over $r$ in the second line in \eqref{LowTComputationGeneralpW} can be evaluated analytically to give 
\begin{align} \label{FiniteS1a}
2 \sin \Big( \frac{\ell_t l (w n - p\pi)}{2\pi L n} \Big) \sin \Big( \frac{\ell_t l p }{2 L n} \Big)  \frac{\sin (\frac{\ell_t l }{2L} )}{\sin (\frac{\ell_t l}{2L n} )}  \;.  
\end{align} 
Thus $	\log \tilde Z_{s}[n]$ is the combination of \eqref{LowTComputationSmallw} and this result. 

The corresponding normalization factor can be obtained by taking $n \to 1$ limit for \eqref{FiniteS1a} and further corrections due to $\tilde \alpha_s$ in \eqref{NormalizationSpin}.
\begin{align}  \label{NormalizationFinp1}
\log \tilde Z_{s,p}[1] &=4 \sum_{l=1}^{\infty} \frac{(-1)^{l-1}}{l} 
\frac{\sinh (\frac{l\beta }{2}) \cos (\frac{l \alpha }{2})}{\cosh ( l\beta ) - \cos (l \alpha)} \bigg\{ \cos \big( \frac{\ell_t l}{2\pi L} w \big)  -1   \Big.  \nonumber \\
& \Big. + 2 \sin \Big( \frac{\ell_t l (w - p\pi)}{2\pi L } \Big) \sin \Big( \frac{\ell_t l p }{2 L } \Big) + \sum_{r=1}^p \frac{2}{n-1} \sin \Big( \frac{\ell_t l ( 2r-1) (n-1)}{2 L n} \Big) \sin \Big( \frac{\ell_t l  }{2 L } \Big) \bigg\}
\;.
\end{align} 
The first term in the second line comes from the $n \to 1$ limit for \eqref{FiniteS1a}. 
The second term comes from the matching condition that the entropies are continuous for each topological transitions, whose sum over $r$ can be done explicitly. 
Thus, combining all together, the R\'enyi entropy is given by 
\begin{align} \label{LowTRenyiFiniteS1a}
\begin{split}
&S_{n,p\leq n}^{w,\mu=0,J=0}
=S_{n}^{w,\mu=0,J=0} \\
&\quad+ 4\sum_{l=1}^{\infty} \frac{(-1)^{l-1}}{l} 
\frac{\sinh (\frac{l\beta }{2}) \cos (\frac{l \alpha }{2})}{\cosh ( l\beta ) - \cos (l \alpha)} \frac{1}{1-n} 
\Bigg\{  
2 \sin \Big( \frac{\ell_t l (w n - p\pi)}{2\pi L n} \Big) \sin \Big( \frac{\ell_t l p }{2 L n} \Big)  \frac{\sin (\frac{\ell_t l }{2L} )}{\sin (\frac{\ell_t l}{2L n} )}  \\
&\left.\quad 
-2 n \sin \Big( \frac{\ell_t l (w - p\pi)}{2\pi L } \Big) \sin \Big( \frac{\ell_t l p }{2 L } \Big) 
\-n \Big(1\- \cos \Big(\frac{\ell_t l p(n-1)}{Ln} \Big)\Big) \frac{\sin ( \frac{\ell_t l }{2 L } )}{ \sin ( \frac{\ell_t l (n-1) }{2 L n} ) }
\right\} \;,
\end{split}
\end{align}
which is valid $\frac{(2p-1)\pi}{n} \leq w < \frac{(2p+1)\pi}{n} $ and $0< p\leq n$. $ S_{n}^{w,\mu=0,J=0}$ is given in \eqref{LowTRenyiSmallW}. When $p=n$, one can get the R\'enyi entropy with $l_k=-1$ for all $k$. 

Next, we consider the case with the number of the topological transitions are bigger than the replica copies, $n<Q \leq 2 n$, so that we have $|l_k|=1,2$ for all $k$. $Q=n+p'=\sum_k |l_k| $. It turns out that the effects of the topological transition to the entropies are similar to the previous cases.  
\begin{align}  \label{LowTComputationGeneralpW2}
\begin{split}
\log \tilde Z_{s}[n] 
&= 4 \sum_{l,m=1}^{\infty}  \frac{(-1)^{l-1}}{l} 
\frac{\cos (\alpha l (m-1/2)) }{e^{l\beta (m-1/2 )}} \Big[ \sum_{k=-\frac{n-1}{2}}^{\frac{n-1}{2}} \cos \Big( \frac{\ell_t l}{L} \big[\frac{k}{n} + \frac{w}{2\pi} \big] \Big) - n   \\
& + \sum_{r=1}^n \bigg\{ \cos \Big( \frac{\ell_t l}{L} \big[\frac{n\-(2r-1)}{2n} + \frac{w}{2\pi} - 1 \big] \Big)  -  \cos \Big( \frac{\ell_t l}{L} \big[\frac{n-(2r-1)}{2n} + \frac{w}{2\pi}  \big] \Big) \Big\} \\
& + \sum_{r'=1}^{p'} \Big\{ \cos \Big( \frac{\ell_t l}{L} \big[\frac{n\-(2r'-1)}{2n} + \frac{w}{2\pi} - 2 \big] \Big)  -  \cos \Big( \frac{\ell_t l}{L} \big[\frac{n-(2r'-1)}{2n} + \frac{w}{2\pi} -1 \big] \Big) \Big\}   \Big] \;. 
\end{split}
\end{align} 
The first line is the same as \eqref{LowTComputationSmallw} and so are the entropies. The second line can be obtained by setting $p=n$ in \eqref{LowTRenyiFiniteS1a}. The third line contains all the new terms due to the topological transitions for $n <Q=n+p' \leq 2n$. For example, $Q=n+1$ case is relevant for $(2n+1) \pi/n \leq w < (2n+3) \pi/n$. The sum over $r$ in the third line in \eqref{LowTComputationGeneralpW2} can be evaluated analytically to give 
\begin{align} \label{FiniteS2a}
2 \sin \Big( \frac{\ell_t l (w n - (2n+p')\pi)}{2\pi L n} \Big) \sin \Big( \frac{\ell_t l p' }{2 L n} \Big)  \frac{\sin (\frac{\ell_t l }{2L} )}{\sin (\frac{\ell_t l}{2L n} )}  \;.  
\end{align} 
Thus $	\log \tilde Z_{s}[n]$ is the combination of \eqref{LowTComputationSmallw}, \eqref{LowTRenyiFiniteS1a} with $p=n$, and this result. Note that \eqref{FiniteS2a} is effectively the same as \eqref{FiniteS1a} because the range of $nw$ is larger than $2n \pi$ that compensates the $2n\pi$ of the contributions $(2n+p')\pi$ due to the number of topological transitions. This tells us that $\tilde \alpha_s$ is repetitive in the sequence for each $N$ in the range $(2Nn+1) \pi/n \leq w < (2Nn+3) \pi/n$.

The corresponding normalization factor is given by
\begin{align}  
\begin{split}
\log \tilde Z_{s}[1] &= \log \tilde Z_{s,p=n}[1] \\
&+ 4 \sum_{l=1}^{\infty} \frac{(-1)^{l-1}}{l} 
\frac{\sinh (\frac{l\beta }{2}) \cos (\frac{l \alpha }{2})}{\cosh ( l\beta ) - \cos (l \alpha)} \bigg\{ 2 \sin \Big( \frac{\ell_t l (w - (2n+p')\pi)}{2\pi L } \Big) \sin \Big( \frac{\ell_t l p' }{2 L } \Big)  \Big.  \\
& \Big.\qquad \qquad  + \sum_{r'=1}^{p'} \frac{2}{n-1} \sin \Big( \frac{\ell_t l ( 2r'-1) (n-1)}{2 L n} \Big) \sin \Big( \frac{\ell_t l }{2 L } \Big) \bigg\}
\;,
\end{split}
\end{align} 
where $\log \tilde Z_{s,p=n}[1]$ contains all the normalization terms for the first two lines of \eqref{LowTComputationGeneralpW2} and is given in \eqref{NormalizationFinp1}.
The last two lines are from the transitions with $l_k=-2,-1$. Then, the R\'enyi entropy is   
\begin{align} \label{LowTRenyiFiniteS2a}
&S_{n,n<Q=n+p'\leq 2n}^{w,\mu=0,J=0}
=S_{n, p=n}^{w,\mu=0,J=0}  \nonumber \\
&\quad+ 4\sum_{l=1}^{\infty} \frac{(-1)^{l-1}}{l} 
\frac{\sinh (\frac{l\beta }{2}) \cos (\frac{l \alpha }{2})}{\cosh ( l\beta ) - \cos (l \alpha)} \frac{1}{1-n} 
\Big\{  
2 \sin \Big( \frac{\ell_t l (w n - (2n+p')\pi)}{2\pi L n} \Big) \sin \Big( \frac{\ell_t l p' }{2 L n} \Big)  \frac{\sin (\frac{\ell_t l }{2L} )}{\sin (\frac{\ell_t l}{2L n} )}  \nonumber \\
&\quad 
-2 n \sin \Big( \frac{\ell_t l (w - (2+p')\pi)}{2\pi L } \Big) \sin \Big( \frac{\ell_t l p' }{2 L } \Big) 
\-n \Big(1\- \cos \Big(\frac{\ell_t l p'(n-1)}{Ln} \Big)\Big) \frac{\sin ( \frac{\ell_t l }{2 L } )}{ \sin ( \frac{\ell_t l (n-1) }{2 L n} ) }
\Big\} \;,
\end{align}
which is valid $\frac{(2Q-1)\pi}{n} \leq w < \frac{(2Q+1)\pi}{n} $ with $Q=n+p'$. $ S_{n,p=n}^{w,\mu=0,J=0}$ is given in \eqref{LowTRenyiFiniteS1a} by setting $p\to n$. 

\begin{figure}[!t]
	\begin{center}
		\includegraphics[width=.62\textwidth]{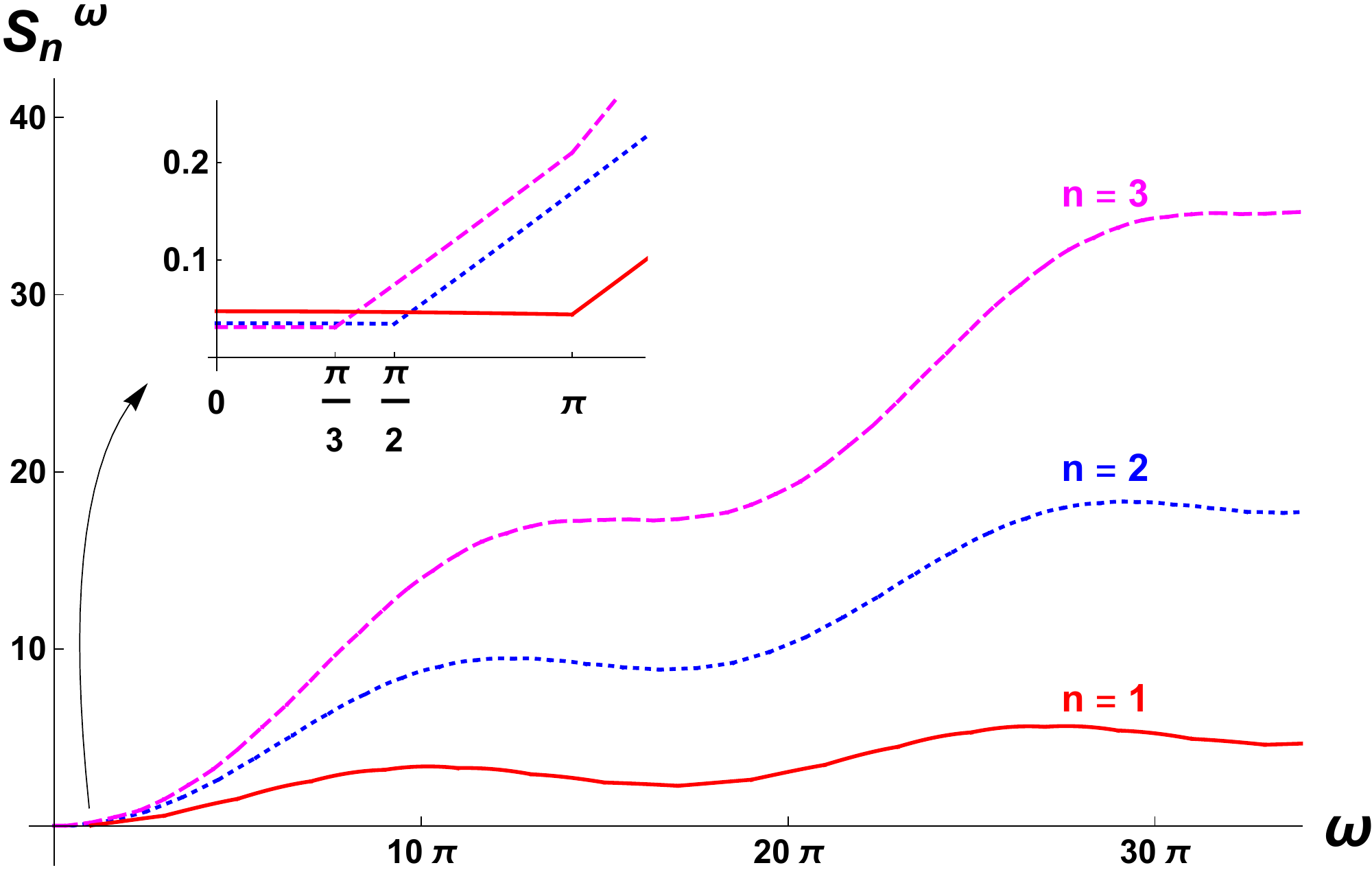} 
		\caption{\footnotesize\small The plot for the R\'enyi entropies \eqref{LowTRenyiFiniteSNa} for $l=1$ and $ 0 \leq w < 31 \pi$ with three different number of $n$, $n=1, 2$, and $3$ represented by red solid, blue dotted and magenta dashed lines, respectively. The inset is for $ 0 \leq w < \pi$ and show the first topological transitions for different $n$.}
		\label{fig:RenyiLowT}
	\end{center}
\end{figure}

Here we can actually consider the general case with more than $Nn$ topological transitions, $|l_k| = N, N-1$. Then, for $Q=Nn+p'$,
\begin{align}  \label{LowTComputationGeneralpWN}
\log \tilde Z_{s}[n] 
&= 4 \sum_{l,m=1}^{\infty}  \frac{(-1)^{l-1}}{l} 
\frac{\cos (\alpha l (m-1/2)) }{e^{l\beta (m-1/2 )}} \bigg[ \sum_{k=-\frac{n-1}{2}}^{\frac{n-1}{2}} \cos \Big( \frac{\ell_t l}{L} \big[\frac{k}{n} + \frac{w}{2\pi} \big] \Big) - n \bigg.    \\
& + \sum_{S=1}^N\sum_{r=1}^n \bigg\{ \cos \Big( \frac{\ell_t l}{L} \big[\frac{n\-(2r-1)}{2n} + \frac{w}{2\pi} \- S \big] \Big)  \-  \cos \Big( \frac{\ell_t l}{L} \big[\frac{n-(2r-1)}{2n} + \frac{w}{2\pi}  \-(S\-1) \big] \Big) \bigg\}  \nonumber \\
& + \sum_{r'=1}^{p'} \bigg\{ \cos \Big( \frac{\ell_t l}{L} \big[\frac{n\-(2r'-1)}{2n} \+ \frac{w}{2\pi} \- (N\+1) \big] \Big)  \-  \cos \Big( \frac{\ell_t l}{L} \big[\frac{n-(2r'-1)}{2n} + \frac{w}{2\pi} \-N \big] \Big) \bigg\}  \bigg.  \bigg] \;. \nonumber
\end{align} 
This holds for $(2Nn+2p-1) \pi/n \leq w < (2Nn+2p+1) \pi/n$. The corresponding normalization factor can be obtained by the iterative process following above. The R\'enyi entropy is given by
\begin{align} \label{LowTRenyiFiniteSNa}
\begin{split}
&S_{n,Nn<Q=Nn+p' \leq (N+1)n}^{w,\mu=0,J=0}= 4\sum_{l=1}^{\infty} \frac{(-1)^{l-1}}{l} 
\frac{\sinh (\frac{l\beta }{2}) \cos (\frac{l \alpha }{2})}{\cosh ( l\beta ) - \cos (l \alpha)} \frac{1}{1-n} 
\Bigg\{  \frac{\sin (\frac{\ell_t l}{2L} )}{\sin (\frac{\ell_t l}{2L n} )}  -n \\
& \quad \quad  - 2n \sin \Big( \frac{\ell_t l (w  - (N+n-1)\pi)}{2\pi L} \Big) \sin \Big( \frac{\ell_t l N }{2 L } \Big)  \frac{\sin (\frac{\ell_t l n}{2L} )}{\sin (\frac{\ell_t l}{2L } )}  \\
&\quad \quad  
+2  \sin \Big( \frac{\ell_t l (w - N\pi)}{2\pi L } \Big) \sin \Big( \frac{\ell_t l N}{2 L } \Big) \frac{\sin (\frac{\ell_t l }{2L} )}{\sin (\frac{\ell_t l}{2L n} )}
\-n N\Big(1\- \cos \Big(\frac{\ell_t l (n-1)}{L} \Big)\Big) \frac{\sin ( \frac{\ell_t l }{2 L } )}{ \sin ( \frac{\ell_t l (n-1) }{2 L n} ) }\\
& \quad \quad  + 2 \sin \Big( \frac{\ell_t l (w n - (2Nn+p')\pi)}{2\pi L n} \Big) \sin \Big( \frac{\ell_t l p' }{2 L n} \Big)  \frac{\sin (\frac{\ell_t l }{2L} )}{\sin (\frac{\ell_t l}{2L n} )}  \\
&\left.\quad \quad  
-2 n \sin \Big( \frac{\ell_t l (w - (2N+p')\pi)}{2\pi L } \Big) \sin \Big( \frac{\ell_t l p' }{2 L } \Big) 
\-n \Big(1\- \cos \Big(\frac{\ell_t l p'(n-1)}{Ln} \Big)\Big) \frac{\sin ( \frac{\ell_t l }{2 L } )}{ \sin ( \frac{\ell_t l (n-1) }{2 L n} ) }
\right\} \;,
\end{split}
\end{align}
which is valid $\frac{(2Q-1)\pi}{n} \leq w < \frac{(2Q+1)\pi}{n} $ with $Q=Nn+p'$. The first line comes from \eqref{LowTComputationSmallw}. The second and third lines are from the contributions of $Nn$ topological transitions. The last two lines are from $Nn+1$ to $ Nn+ p'$ transitions. 

\begin{figure}[!b]
	\begin{center}
		\includegraphics[width=.66\textwidth]{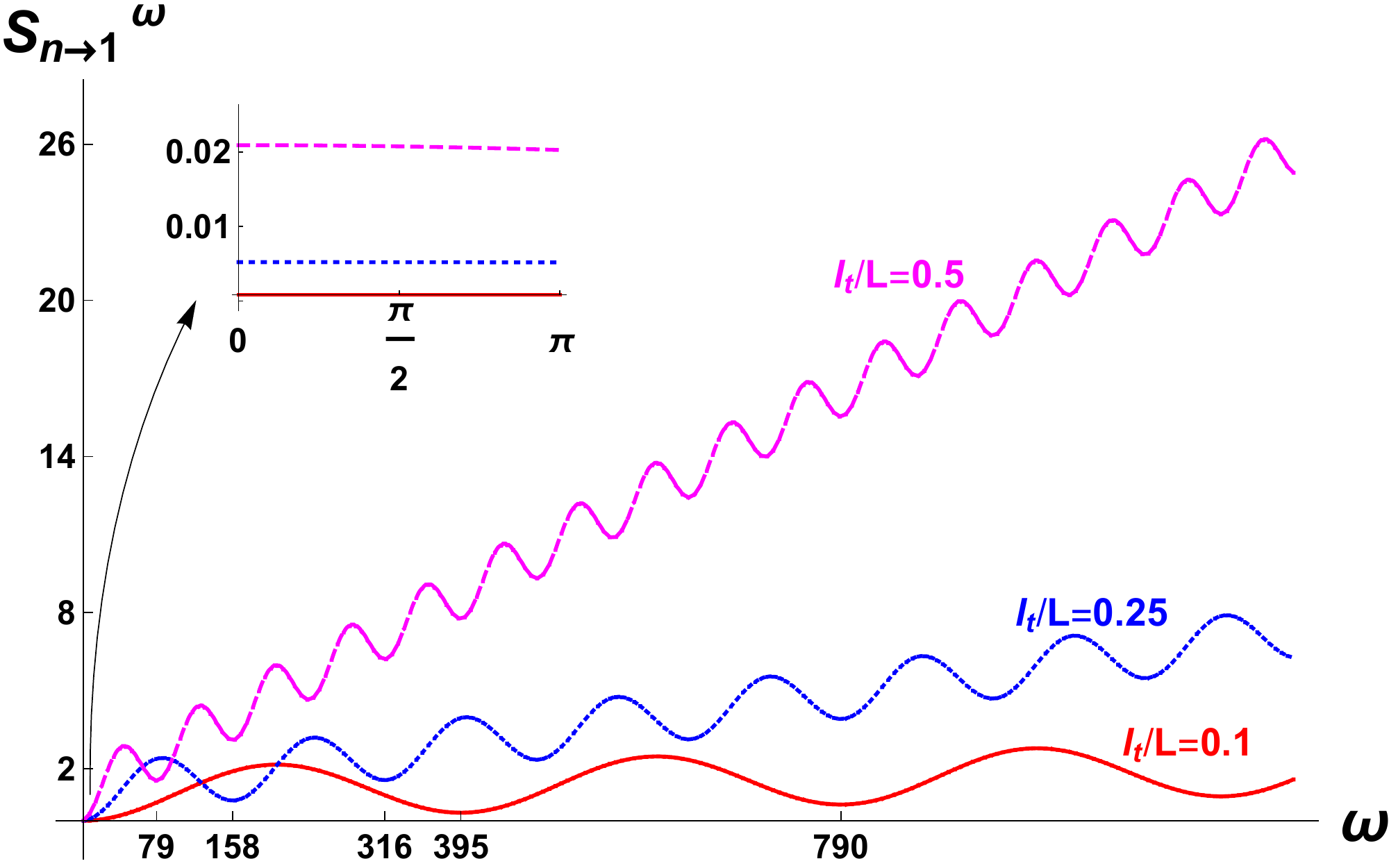} 
		\caption{\footnotesize\small The plot for entanglement entropy \eqref{LowTEEFiniteSNa} for $l=1$ and $ 0 \leq w < 401 \pi$ with three different sub-system sizes, $\ell_t/L=0.1, 0.25$, and $0.5$ represented by red solid, blue dotted and magenta dashed lines, respectively. The corresponding inset is for $ 0 \leq w < \pi$. }
		\label{fig:EELowT}
	\end{center}
\end{figure}

When taking $n\to 1$ limit, $p'$ in \eqref{LowTRenyiFiniteSNa} becomes a worth of the topological transitions with the number of replica copies. The range of $w$ becomes $(2p-1)\pi \leq w <  (2p+1)\pi$ with $p=N+p'$. Thus, it is convenient to take the entanglement entropy limit for the first three lines in \eqref{LowTRenyiFiniteSNa} 
\begin{align} \label{LowTEEFiniteSNa}
\begin{split}
&S_{n\to 1}^{w,\mu=0,J=0}= 4\sum_{l=1}^{\infty} \frac{(-1)^{l-1}}{l} 
\frac{\sinh (\frac{l\beta }{2}) \cos (\frac{l \alpha }{2})}{\cosh ( l\beta ) - \cos (l \alpha)} 
\Bigg\{ \cos \Big( \frac{\ell_t l}{2\pi L} w \Big) \Big[ 1 - \frac{\ell_t l}{2L} \cot (\frac{\ell_t l}{2L} ) \Big]  \\
& \qquad \quad  +2 \sin \Big( \frac{\ell_t l p }{2 L } \Big)   \sin \Big( \frac{\ell_t l (w - p\pi)}{2\pi L } \Big) \Big[ 1 - \frac{\ell_t l}{2L} \cot (\frac{\ell_t l (w - p\pi)}{2L} ) \Big]
+ p \frac{\ell_t l}{L} \sin ( \frac{\ell_t l }{2 L } ) \Bigg\} \;.
\end{split}
\end{align}
Entanglement entropy is periodic function with a slow linear growth as a function of $w$, more precisely as a function of $p$, due to the last term. The R\'enyi and entanglement entropies given in \eqref{LowTRenyiFiniteSNa} and \eqref{LowTEEFiniteSNa} are depicted in figure \ref{fig:RenyiLowT}. One can see how the entanglement entropy can be obtained pictorially from the figure. We also present this result \eqref{LowTEEFiniteSNa} in the figure \ref{fig:EELowT} for the three different sub-system sizes described in the units of $\frac{\sinh (\frac{l\beta }{2}) \cos (\frac{l \alpha }{2})}{\cosh ( l\beta ) - \cos (l \alpha)} $ for $l=1$.  Their corresponding periodicities are $ 4\pi^2/(\ell_t / L )= 395, 158$, and $79 $ for $ \ell_t / L =0.1, 0.25$, and $0.5$, respectively. It turns out that the first line of \eqref{LowTEEFiniteSNa} contribute minimally.

We note that the Wilson loop winding sector restricts the range of $w$. If one starts with Wilson loops sector with $\gamma=2$, then the corresponding range is $-2\pi \leq w < 2\pi $. Then, one needs to exclude the region $w>2\pi$ in figures \ref{fig:RenyiLowT} and \ref{fig:EELowT}. We mention that entanglement entropy is continuous across the Wilson loops sectors.  

\subsubsection{Large radius limit: with full generality} 

The large radius limit can be conveniently evaluated by changing the theta function in a slightly different form. 
$\vartheta_3 (z|\tau) = \prod_{m=1}^\infty (1 - q^m)(1 + q^{2m-1} + 2 \cos (2\pi z) q^{m-1/2})$, where $q=e^{2\pi i \tau}$. Instead of expanding $q$, we expand a small factor $\delta z$ in $z_1 = z_2 + \delta z$ with $ \delta z \propto \ell_t/L$. Thus, 
\begin{align} 
	&\Big| \frac{(1 - q^m)(1 + y_1 q^{m-1/2})(1 + y_1^{-1} q^{m-1/2})}{(1 - q^m)(1 + y_2 q^{m-1/2})(1 + y_2^{-1} q^{m-1/2})} \Big|^2  
	= \Big| \frac{1 + q^{2m-1} + 2 \cos (2\pi z_1) q^{m-1/2}}{1 + q^{2m-1} + 2 \cos (2\pi z_2) q^{m-1/2}} \Big|^2  \nonumber \\
	&= \Big| 1 + \mathcal O \left(\frac{\ell_t^2}{4\pi^2 L^2} \right) \Big|^2 
	\;. 
\end{align} 
This shows that the $w$ dependence of the spin dependent entropies vanishes as fast as $\frac{\ell_t^2}{4\pi^2 L^2} $, meaning that $w$ dependence vanishes in infinite space. 

Actual computation of \eqref{TopologicalJMUW22} for large radius limit can be done by observing $ z_1 = \delta z=  \frac{\ell_t \alpha_{w,k} }{2\pi L} =  \frac{\ell_t}{2\pi L} \left[\frac{k}{n} + \frac{w}{2\pi} + l_k \right] \ll 1, z_2=0$. One can easily see the same summation factor that is in \eqref{SpinIndGenComputationP} and the related discussions.  
\begin{align}
\begin{split}	
	S_{n, L\to \infty}^{w, \mu =J =0} &= \frac{-1}{1 -n} \Big[  \sum_{k =-\frac{n-1}{2}}^{\frac{n-1}{2}} \alpha_{w,k}^2 - n \alpha_{w,0}^2 \Big]  \\
	& \times  \frac{\ell_t^2}{4 L^2} \Big[ 1 + \tanh^2 \left(\frac{\beta\mu}{2} \right) +\sum_{r=1}^p \frac{4+4 \cosh(\beta \mu) \cosh(\beta r)}{\cosh (\beta \mu) + \cosh (\beta r)} \Big]    
	\;.
\end{split}	
\end{align}
Here we observe that the front factor is the same as \eqref{EEIndSpinS}. Thus, 
\begin{align} \label{LargeREEGeneral}
	S_{n, L\to \infty}^{w,\mu=J=0} &= \Big( \frac{n+1}{12 n} + Q\frac{w}{\pi} - \frac{Q^2}{n} \Big) ~\frac{\ell_t^2}{4 L^2} \left[ 1 + \tanh^2 \left(\frac{\beta\mu}{2} \right) +\sum_{r=1}^p \frac{4+4 \cosh(\beta \mu) \cosh(\beta r)}{\cosh (\beta \mu) + \cosh (\beta r)} \right] 
	\;,
\end{align}
where the result is valid for $(2Q-1)\pi/n \leq w < (2Q+1)\pi/n$. Thus we confirm the spin structure dependent entropies vanish as fast as $\mathcal O(\frac{\ell_t^2}{4 L^2} ) $ at the large radius limit $L\to \infty$. We note that this result for the large radius limit applies for all the spin dependent entropies regardless of the parameters, $w, \mu, J, a$ and $b$. Thus, we do not consider the large radius limit for the other cases we consider. 

Importantly, there exist finite and dominant contributions in the large radius limit from the spin structure independent entropies. Those contributions in equations \eqref{RenyiIndSpinSWithQ} and \eqref{EEIndSpinSWithQ} turn out to be exact. This is mentioned in the previous section and also given in \cite{Kim:2017xoh}.

\subsection{Entanglement entropy $S_n^{w,\mu,J}$ with $\mu, J$. } \label{sec:GeneralLowTEEMuJWGan}

Here we present the physical effects of the Wilson loops in the presence of the chemical potential and current of the entropies. In particular, we focus on the {\it anti-periodic fermions} in the low temperature limit, $\beta=2\pi \tau_2 \to \infty$. Thus $a+J=b=1/2$. We start with computing  
\begin{align}
	\log \tilde Z_{s}[n] = \sum_{k=-{(n-1)}/{2}}^{{(n-1)}/{2}} \log \big| \frac{\vartheta_3 (\frac{\ell_t}{2\pi L} \left[\frac{k}{n} + \frac{w}{2\pi} + l_k \right] + \tau_1 J + i \tau_2 \mu|\tau)}{\vartheta_3 (\tau_1 J + i \tau_2 \mu|\tau)} \big|^2 \;,
\end{align}	 
in \eqref{TopologicalEEFormulaJmu1}. In the low temperature limit, one expands the Jacobi theta function and factorize various contributions. The case for $w=0$, and thus $l_k=0$, is explicitly computed in \cite{Kim:2017ghc}. 

Before presenting the results for the chemical potential and current dependences in the low temperature limit $\beta \to \infty$, we note that there is an important subtlety that has been sorted out previously \cite{Kim:2017xoh}\cite{Kim:2017ghc}. When $\mu =1/2, 1, 3/2, \cdots$ one of the energy levels of the Dirac fermions, we need to be careful of the factor $e^{\beta \mu}$ and need to take care of a finite contribution that comes from $ \beta \to \infty, \mu -1/2 \to 0$. Similarly, the factor $e^{-\beta\mu}$ needs care for $\mu=-1/2, -1, -3/2, \cdots$. For more details, refer to \cite{Kim:2017xoh}\cite{Kim:2017ghc}. We carefully revisited this subtlety in the presence of the Wilson loop parameter $w$ to conclude that these two do not interfere with each other both in the low temperature and large radius limits. Thus we present the effects of the Wilson loops parameter $w$ without this subtle point. Here we focus on the entropies that are only valid for $ -1/2 < \mu < 1/2$. Extending the analysis for other values and including these special values of $\mu$ is straightforward.  \\

\noindent {\it $\square$ R\'enyi and entanglement entropies $S_n^{w,J,\mu}$ for $ w < \pi/n$. } 

The computations for the entropies with the topological Wilson loops $w$ in the presence of $J$ and $\mu$ are a little bit complicated. We consider the case $w<\pi/n$ first with all the vanishing $\ell_k$'s, $l_{(n-1)/2} = \cdots =l_{-(n-1)/2} =0$. For low temperature limit, we expand the Jacobi theta functions and collect the terms to have the following cosine form. 
\begin{align}  \label{LogZnMuJ}
\begin{split}
\log \tilde Z_{s}[n] 
&=\!\! \sum_{l,m=1}^{\infty}  \frac{2 (-1)^{l-1} }{l e^{l\beta (m-\frac{1}{2} )}} 
\bigg[  e^{l\beta \mu} \!\! \!\!\sum_{k=-{(n-1)}/{2}}^{{(n-1)}/{2}} \!\!\!  \Big\{ \cos \big( \frac{\ell_t l}{L} \alpha_{w,k} \+ \alpha l J_{-m}  \big) \- \cos \big( \alpha l J_{-m}  \big)  \Big\}    \\
&\quad\quad  +  e^{-l\beta \mu} \!\! \!\!\sum_{k=-{(n-1)}/{2}}^{{(n-1)}/{2}} \!\!\!  \Big\{ \cos \big( \frac{\ell_t l}{L} \alpha_{w,k} \+ \alpha l J_{m}  \big) \- \cos \big( \alpha l J_{m}  \big)  \Big\} \bigg] \;.
\end{split}
\end{align} 
The expression becomes long with the contribution of current source $J$, thus we abbreviate $J_{m} \equiv  J\+m\-\frac{1}{2} $ and $J_{-m} \equiv  J\-m\+\frac{1}{2} $.
The summation over the index $k$ in the first line can be done analytically to give 
\begin{align}\label{SumOverK22}
\cos \left( \frac{l \ell_t }{ L} \frac{w}{\pi} +  \alpha l J_{-m} \right) \frac{\sin \left(\frac{l \ell_t}{2 L}\right)}{ \sin \left(\frac{l \ell_t}{2 L n}\right)}  
-n \cos \left(\alpha l J_{-m}  \right)
\end{align}
Similar expression can be obtained for the summation in the second line by replacing $J_{-m} \to J_{m}  $. 

We choose the normalization so that the R\'enyi entropy is continuous across the topological sectors and has a smooth entanglement entropy limit for $n \to 1$. These two conditions can be met with the following choice. 
\begin{align} \label{NormalizationSpinMuJ}
\log \tilde Z_{s}[1] = \log \left( \lim_{n\to 1} \tilde Z_{s}[n] \right) + (n-1) \log \tilde \alpha_{s,\mu,J}  \;.
\end{align} 
For $w< \pi/n$, $l_k=0$ for all $k$. We choose $ \log \tilde \alpha_{s,\mu,J} =0 $. Thus the R\'enyi entropy has the form. 
\begin{align} \label{RenyiWJMu0}
\begin{split}	
S_{n}^{w<\pi/n,\mu,J}
&= 2\sum_{l,m=1}^{\infty} \frac{(-1)^{l-1} }{l e^{\beta l (m-\frac{1}{2})}} 
 \frac{1}{1-n} \big[ \frac{\sin \big(\frac{l \ell_t}{2 L}\big)}{ \sin \big(\frac{l \ell_t}{2 L n}\big)}  -n \big]    \\
&\times  \bigg\{ e^{\beta\mu l} \cos \big( \alpha l J_{-m}  \+ \frac{\ell_t l}{2\pi L} w\big)  + e^{-\beta\mu l} \cos \big( \alpha l J_{m}  \+ \frac{\ell_t l}{2\pi L} w\big)  \bigg\} \;. 
\end{split}
\end{align}
One can check that this result is entirely finite size effect. For $\ell_t/L \to 0$, the entropies vanish as $ \mathcal O(\ell_t/L)^2$. The same is true for the entanglement entropy. 

Finally, the entanglement entropy has the form
\begin{align} \label{EEWJMu0}
\begin{split}	
S_{n\to 1}^{w<\pi,\mu,J}
&= 2\sum_{l,m=1}^{\infty} \frac{(-1)^{l-1} }{l e^{\beta l (m-\frac{1}{2})}}  \big[ 1 - \frac{\ell_t l}{2L} \cot (\frac{\ell_t l}{2L} ) \big]   \\
& \times \bigg\{ e^{\beta\mu l} \cos \big( \alpha l J_{-m}  \+ \frac{\ell_t l}{2\pi L} w\big) 
+ e^{-\beta\mu l} \cos \big( \alpha l J_{m}  \+ \frac{\ell_t l}{2\pi L} w\big)  \bigg\} \;.  
\end{split}
\end{align}
The entanglement entropy is a periodic function of $J$ with periodicity $ J = 2\pi/\alpha l$. 
The result simplifies when one of the torus modulus parameter $\alpha=2\pi \tau_1$ vanishes. Then 
\begin{align} \label{EEMuJW11}
S_{n=1,\alpha=0}^{w<\pi,\mu,J}
&= 4\sum_{l,m=1}^{\infty} \frac{(-1)^{l-1} }{l e^{\beta l (m-\frac{1}{2})}}  \left[ 1 - \frac{\ell_t l}{2L} \cot (\frac{\ell_t l}{2L} ) \right] \cos \left(  \frac{\ell_t l}{2\pi L} w\right) \cosh (\beta\mu l) \;. 
\end{align}
Thus the entropies have periodic dependence on the Wilson loop parameter $w$. \\

\noindent {\it $\square$ R\'enyi and entanglement entropies $S_n^{w,J,\mu}$ for $ (2Q-1) \pi/n \leq w < (2Q+1) \pi/n$. } 

This section is similar to the previous section with additional dependence of $J$ in the low temperature limit. The topological transitions have the same pattern described in the table \ref{table:Transitions}, and the procedure to take care of the normalization factors are the same as done in the previous section \S \ref{sec:FiniteEEMuJZero}. 

To see the differences compared to the previous section, we consider $Q=p \leq n$, so that $l_k=-1, 0$ for all $k$ with a large and finite winding sector. Focusing only with the part proportional to $e^{l\beta\mu}$, we see that the following contributions are added to the first line in \eqref{LogZnMuJ}.
\begin{align}  \label{LowTComputationGeneralpWJ}
 \sum_{r=1}^p \bigg\{ \cos \Big( \frac{\ell_t l}{L} \big[\frac{n\-(2r\-1)}{2n} \+ \frac{w}{2\pi} \- 1 \big] \+ \alpha l J_{-m} \Big)  \-  \cos \Big( \frac{\ell_t l}{L} \big[\frac{n\-(2r\-1)}{2n} \+ \frac{w}{2\pi}  \big]+ \alpha l J_{-m}  \Big) \bigg\}  \;. 
\end{align} 
The new contribution $\alpha l J_{-m}$ appears consistently and straightforward to take care of. The sum over $r$ can be evaluated analytically.
\begin{align} \label{FiniteS1aJ}
2 \sin \Big( \frac{\ell_t l (w n - p\pi)}{2\pi L n} + \alpha l J_{-m} \Big) \sin \Big( \frac{\ell_t l p }{2 L n} \Big)  \frac{\sin (\frac{\ell_t l }{2L} )}{\sin (\frac{\ell_t l}{2L n} )}  \;.  
\end{align} 

We have three contributions for the normalization factor $\log \tilde Z_{s}[1]$. Two contributions come from \eqref{SumOverK22} and \eqref{FiniteS1aJ} (after taking $n\to 1$ limit) for smooth entanglement entropy limit. The last one  
\begin{align}
\log \tilde \alpha_{s,\mu,J} &= \sum_{r=1}^{p} \frac{2\sin ( \frac{\ell_t l }{2 L } ) }{n(n-1)} \bigg\{n \sin \Big( \frac{\ell_t l ( 2r-1) (n-1)}{2 L n} \- \alpha l J_{-m} \Big) \+ \sin \Big(  \alpha l J_{-m} \Big)  \bigg\}
\end{align} 
is necessary to have the continuous entropies across the topological transitions. One can check this reduces to the last contribution in \eqref{NormalizationFinp1} for $\alpha J_{-m} \to 0$. Similar to the previous case, $\log \tilde \alpha_{s,\mu,J}$ contributes to each topological transition. This tells us that we can build up the R\'enyi entropy based on the previous section. 

Putting all together, we get the R\'enyi entropy for the general case. 
\begin{align} \label{LowTRenyiFiniteSNaJ}
\begin{split}
&S_{n,Nn<Q=Nn+p' \leq (N+1)n}^{w,\mu,J}= \sum_{l,m=1}^{\infty} \frac{(-1)^{l-1} }{l e^{\beta l (m-\frac{1}{2})}} 
\frac{2}{1-n}  e^{\beta\mu l} 
\Bigg\{ \cos \big( \alpha l J_{-m}  \+ \frac{\ell_t l}{2\pi L} w\big)  \Big[ \frac{\sin (\frac{\ell_t l}{2L} )}{\sin (\frac{\ell_t l}{2L n} )}  \-n \Big] \\
& \quad \quad  \qquad \qquad  \- 2n \sin \Big( \frac{\ell_t l (w  \- (N\+n\-1)\pi)}{2\pi L} \+\alpha l J_{-m} \Big) \sin \Big( \frac{\ell_t l N }{2 L } \Big)  \frac{\sin (\frac{\ell_t l n}{2L} )}{\sin (\frac{\ell_t l}{2L } )}  \\
&\quad \quad   \qquad \qquad
\+2  \sin \Big( \frac{\ell_t l (w \- N\pi)}{2\pi L } \+\alpha l J_{-m}\Big) \sin \Big( \frac{\ell_t l N}{2 L } \Big) \frac{\sin (\frac{\ell_t l }{2L} )}{\sin (\frac{\ell_t l}{2L n} )} 
\-2 N n \sin(\alpha l J_{-m}) \sin (\frac{\ell_t l }{2L} ) \\
&\quad\quad  \qquad \qquad
\-n N\Big(\cos (\alpha l J_{-m}) \- \cos \Big(\frac{\ell_t l (n\-1)}{L} \- \alpha l J_{-m}\Big)\Big) \frac{\sin ( \frac{\ell_t l }{2 L } )}{ \sin ( \frac{\ell_t l (n\-1) }{2 L n} ) }\\
& \quad \quad  \qquad \qquad 
\+ 2 \sin \Big( \frac{\ell_t l (w n \- (2Nn\+p')\pi)}{2\pi L n} \+\alpha l J_{-m} \Big) \sin \Big( \frac{\ell_t l p' }{2 L n} \Big)  \frac{\sin (\frac{\ell_t l }{2L} )}{\sin (\frac{\ell_t l}{2L n} )}  \\
&\quad \quad   \qquad \qquad
\-2 n \sin \Big( \frac{\ell_t l (w \- (2N\+p')\pi)}{2\pi L } +\alpha l J_{-m}\Big) \sin \Big( \frac{\ell_t l p' }{2 L } \Big)  \-2p' \sin(\alpha l J_{-m}) \sin (\frac{\ell_t l }{2L} ) \\
&\quad \quad   \qquad \qquad
\-n \Big(\cos (\alpha l J_{-m}) \- \cos \Big(\frac{\ell_t l p'(n-1)}{Ln} \-\alpha l J_{-m} \Big)\Big) \frac{\sin ( \frac{\ell_t l }{2 L } )}{ \sin ( \frac{\ell_t l (n\-1) }{2 L n} ) }
\Bigg\} \\
&\quad \quad   \qquad \qquad \qquad\qquad +  \text{Same terms with} \quad \Big\{ e^{\beta\mu l} \to e^{-\beta\mu l} \Big\} \quad \&  \quad \Big\{ \alpha l J_{-m}  \to \alpha l J_{m}  \Big\}  \;,
\end{split}
\end{align}
which is valid $\frac{(2Q-1)\pi}{n} \leq w < \frac{(2Q+1)\pi}{n} $ with $Q=Nn+p'$. The first line comes from \eqref{RenyiWJMu0}. The next three lines are from the contributions of $Nn$ topological transitions. The last three lines are from $Nn+1$ to $ Nn+ p'$ transitions.  

When taking $n\to 1$ limit, $p'$ in \eqref{LowTRenyiFiniteSNa} becomes a worth of the topological transitions with the number of replica copies. The range of $w$ becomes $(2Q-1)\pi \leq w <  (2Q+1)\pi$ with $p=N+p'$. Thus, it is convenient to take the entanglement entropy limit for the first three lines in \eqref{LowTRenyiFiniteSNa} 
\begin{align} \label{LowTEEFiniteSNaJ}
\begin{split}
&S_{n\to 1}^{w,\mu,J}= 2 \sum_{l,m=1}^{\infty} \frac{(-1)^{l-1} }{l e^{\beta l (m-\frac{1}{2})}} e^{\beta\mu l} 
\bigg\{ \cos \Big(\alpha l J_{-m} \+ \frac{\ell_t l}{2\pi L} w \Big) \Big[ 1 \- \frac{\ell_t l}{2L} \cot (\frac{\ell_t l}{2L} ) \Big]  \\
& \qquad \quad  \-2 p \sin (\alpha l J_{-m}) \sin \Big( \frac{\ell_t l }{2 L } \Big)  \Big[ 1 \- \frac{\ell_t l}{2L} \cot (\alpha l J_{-m}) \Big] \\
& \qquad \quad  \+ 2  \sin \Big(\alpha l J_{-m} \+  \frac{\ell_t l (w\-p\pi) }{2\pi L } \Big) \sin \Big( \frac{\ell_t l p}{2 L } \Big)  \Big[ 1 \- \frac{\ell_t l}{2L} \cot \Big(\alpha l J_{-m} \+  \frac{\ell_t l (w\-p\pi) }{2\pi L } \Big) \Big] \bigg\}  \\
&\quad \qquad \qquad +  \text{Same terms with} \quad \Big\{ e^{\beta\mu l} \to e^{-\beta\mu l} \Big\} \quad \&  \quad \Big\{ \alpha l J_{-m}  \to \alpha l J_{m}  \Big\}   \;.
\end{split}
\end{align}
This reduces to \eqref{LowTEEFiniteSNa} when $\alpha=J=\mu=0$. 

\begin{figure}[!t]
	\begin{center}
		\includegraphics[width=.6\textwidth]{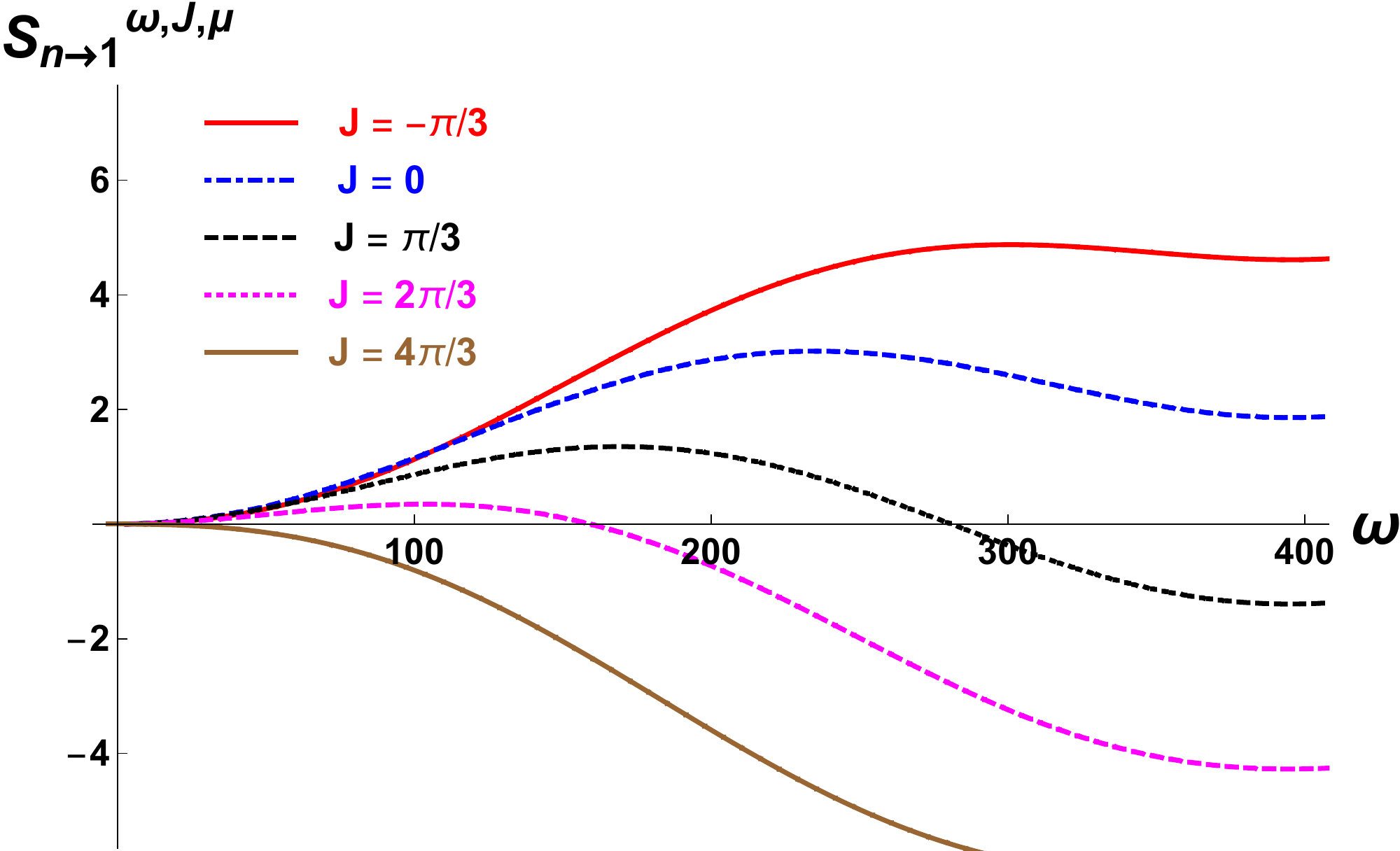} 
		\caption{\footnotesize\small The entanglement entropies \eqref{LowTEEFiniteSNaJ} with at least $80$ topological transitions as $ 0 \leq w < 160\pi$ for $l=1, m=1$ in the unit of $e^{\beta(\mu -1/2)}$ and for an interval $\ell_t = 0.1 L$ for $\alpha=2\pi \tau_1 = 0.5$. As we increase $J$, the entropy becomes negative and returned to the original configuration with the periodicity $\Delta J=2\pi /\alpha=4\pi$.  }
		\label{fig:EELowTGeneralJ}
	\end{center}
\end{figure}

We present this result in the figure \ref{fig:EELowTGeneralJ} for the sub-system sizes $l_t/L=0.1$ and $\alpha=0.5$ for $l=1$. Entanglement entropy is a periodic function as functions of $J$ and $w$ along with a linear growth as a function of $w$, more precisely as a function of $p$, due to the second term. The periodicity in $w$ space is  are $ \Delta w= 4\pi^2/(\ell_t / L )= 395$ for $ \ell_t / L =0.1$. The entropy is also periodic in $J$ with the periodicity $\Delta J= 2\pi/\alpha =4\pi$. As one changes $J$, the entropy move up and down periodically as in the figure \ref{fig:EELowTGeneralJ}.   

The entanglement entropy for the {\it periodic fermion} in the low temperature limit can be accordingly evaluated, and the final result is the same as \eqref{LowTEEFiniteSNaJ} after replacing $m-\frac{1}{2}$ to $m$. Thus the periodic fermion behave similarly as the anti-periodic fermion for the Wilson lop parameter $w$. 

\section{High temperature limit} \label{HighTLimit}

In this section, we consider the high temperature limit of the entropies. As mentioned at the very end of the section \S \ref{sec:SSIndepN}, spin structure independent entropies have interesting dependences on $w$ similar to \eqref{RenyiIndSpinSWithQ} and \eqref{EEIndSpinSWithQ}. Here we focus on the spin structure dependent entropies. For the anti-periodic fermions, we can use the S-duality formula for the theta-function, $ \vartheta_3 (z|\tau) = (-i \tau)^{-1/2} e^{-\pi i z^2/\tau} \theta_3(z/\tau|-1/\tau) $. 

For the general spin dependent entropies, the formula \eqref{TopologicalEEFormulaJmu1} can be rewritten as 
\begin{align}
S_{n,H}^{w,\mu,J} = \frac{1}{1- n} \Big[ \log \tilde Z^H_{s}[n]  -  n \log \tilde Z^H_{s}[1] \Big] \;,
\end{align}
where 
\begin{align} \label{TopologicalEEFormulaJmuHighT}
	\tilde Z^H_{s}[n] &=\!\!\!\! \prod_{k=-{(n-1)}/{2}}^{{(n-1)}/{2}} \Big| \frac{e^{-\frac{i\pi}{\tau} (\frac{\ell_t \alpha_{w,k} }{2\pi L} + \tau_1 J + i \tau_2 \mu)^2} }{e^{-\frac{i\pi}{\tau} ( \tau_1 J + i \tau_2 \mu)^2}}  \frac{\vartheta [\substack{1/2-b \\ 1/2-a-J }](\frac{\ell_t}{2\pi L} \frac{\alpha_{w,k}}{\tau} \+ \frac{\tau_1 J + i \tau_2 \mu}{\tau}|\frac{\-1}{\tau})}{\vartheta [\substack{1/2-b \\ 1/2-a-J }](\frac{\tau_1 J + i \tau_2 \mu}{\tau}|\frac{\-1}{\tau})} \Big|^2 \;, 
\end{align}
and 
\begin{align}	
	\log \tilde Z^H_{s}[1] &= \log \big( \lim_{n\to 1} \tilde Z^H_{s}[n] \big) + (n-1) \log \alpha^H_{s,\mu,J} \;,   \label{JmuHighTNormalization}
\end{align} 
where we take into account the possibility to use different normalization $\log \tilde Z^H_{s}[1]$ for different topological sectors similar to \eqref{Normalization}. We previously noticed that there are two different limits in the high temperature entropies depending on $\alpha=2\pi \tau_1$ \cite{Kim:2017xoh}\cite{Kim:2017ghc}. The two high temperature limits with $\alpha=0$ and $\alpha \neq 0$ are qualitatively different. We consider a simpler case $\alpha=0$ first. \\

\noindent {\it $\square$ Entanglement entropy $S_{n,H}^{w,\mu,J}$ with $\alpha=0$. } 

We consider the anti-periodic fermions for $\alpha=2\pi \tau_1=0$, $ \tau = i \tau_2 = i \frac{2\pi}{\beta}$. To see the physical effects of the Wilson loops parameter $w$ clearly, we further simplify the situation by taking $\mu =J =0$. We observe the first two exponential factors in \eqref{TopologicalEEFormulaJmuHighT} lead similar computations as in \S \ref{sec:SSIndepN}. For the general case with $p$ transitions, the contribution with imaginary part automatically projected out, and 
\begin{align} 
\sum_{k=-(n-1)/2}^{(n-1)/2} \log \big| e^{-\frac{2\pi^2}{\beta} (\frac{\ell_t  }{2\pi L} \big[\frac{k}{n} + \frac{w}{2\pi} + l_k  \big] )^2}  \big|^2 
= - \sum_{k=-(n-1)/2}^{(n-1)/2} \frac{\ell_t^2}{\beta L^2} \big( \frac{k}{n} + \frac{w}{2\pi} + l_k \big)^2 \;. 
\end{align} 
The right hand side is already evaluated with care in \S \ref{sec:SSIndepN} after taking into account various $l_k$'s depending on the topological sectors based on the pattern outlined in the table \ref{table:Transitions}. We note a relative $-$ sign compared to the low temperature case. Thus, after taking care of all the $l_k$'s, we have the R\'enyi entropy for general $Q = \sum_k |l_k| $ as \eqref{RenyiIndSpinSWithQ} or \eqref{EEIndSpinSWithP}
\begin{align} \label{HighTDominant}
\big( \frac{n+1}{12 n} + Q\frac{w}{\pi} - Q^2 \frac{1}{n} \big)\frac{\ell_t^2}{\beta L^2}  \;, 
\end{align} 
which turns out to be the dominant contribution for $\beta \to 0$ limit. 

We compute the theta function contributions in \eqref{TopologicalEEFormulaJmuHighT} (with $\mu\=J\=0$) following \S \ref{sec:FiniteEEMuJZero}. The Jacobi theta function $\vartheta_3 (z|\tau) = \prod_{m=1}^\infty (1 - q^m)(1 + y q^{m-1/2})(1 + y^{-1} q^{m-1/2})$ for large temperature limit $\beta \to 0$ has a well defined expansion with $ q=e^{-4\pi^2/\beta}$ along with $y=e^{2\pi \alpha_{w,k}/\beta } $ in the numerator or $y=1$ in the denominator. After the sum over the indices $k$ and $m$. We get 
\begin{align}  \label{HighTComputationFactor2}
\begin{split}
	&\sum_{k=-(n-1)/2}^{(n-1)/2} \!\! \log \Big| \frac{\vartheta_3 (\frac{\ell_t}{2\pi L} \frac{\alpha_{w,k}}{\tau}|\frac{\-1}{\tau})}{\vartheta_3(0|\frac{\-1}{\tau})} \Big|^2 
	=\sum_{l=1}^{\infty} \frac{2(-1)^{l-1}}{l \sinh ( \frac{2\pi^2}{\beta} l ) } \Big\{ \cosh \big( \frac{\ell_t l}{\beta L} w\big) \frac{\sinh (\frac{\pi \ell_t l}{\beta L} )}{\sinh (\frac{\pi \ell_t l}{\beta L  n} )}  -n \big. \\
	& \qquad\qquad  \+ \sum_{r=1}^p  \Big [ \cosh \Big( \frac{2\pi \ell_t l}{\beta L} \big[\frac{n\-(2r\-1)}{2n} + \frac{w}{2\pi} \-1 \big] \Big) \- \cosh \Big( \frac{2\pi \ell_t l}{\beta L} \big[\frac{n\-(2r\-1)}{2n} \+ \frac{w}{2\pi} \big] \Big) \Big] \Big\} 
	\;, 
\end{split}	
\end{align} 
where the expression is valid for $Q=p\leq n$ that has less or equal to $n$ topological transitions. This is similar to \eqref{LowTComputationGeneralpW} for the low temperature limit. 

We are ready to take into account the normalization factor. Following \eqref{JmuHighTNormalization}, the normalization factor $\log \tilde Z^H_{s}[1] $ can be computed iteratively. For $p \leq n$, 
\begin{align}  
\begin{split}
\log \big( \lim_{n\to 1} \tilde Z^H_{s}[n] \big) &= \sum_{l} \cdots \Big\{  \cosh \big( \frac{\ell_t l}{\beta L} w\big)  - 1 - 2 \sinh \big( \frac{l \ell_t \pi p}{L\beta} \big) \sinh \big( \frac{l \ell_t (w-p \pi)}{L\beta } \big) \Big\}\;, \\
\log \alpha^H_{s,\mu=J=0} &=  \sum_{l} \cdots \times \frac{1}{1-n} \Big\{  2 \sinh \big( \frac{l \ell_t \pi}{L\beta} \big)  \sinh \big( \frac{(2p-1) l \ell_t (n-1) \pi}{L\beta n} \big) \Big\}
\;.  
\end{split}
\end{align} 
We note that, as the phase transition continues beyond $Q>n$, the same normalization $\log \alpha^H_{s,\mu=J=0} $ continues to help the R\'enyi entropy be continuous.

Putting them together, one can get the R\'enyi entropy. The derivation is parallel to the low temperature case done in \S \ref{sec:FiniteEEMuJZero}. Here, we present the entanglement entropy with $p$ transitions for the anti-periodic fermions. 
\begin{align}   \label{HighTEEWithPTerms1}
\begin{split}
S_{n \to 1,p,H}^{w,\mu=J=0} &=\frac{\ell_t^2}{\beta L^2} \big(  \frac{1}{6} + Q\frac{w}{\pi} - Q^2  \big) \\
&+ \sum_{l=1}^{\infty} \frac{2(-1)^{l-1}}{l \sinh ( \frac{2\pi^2}{\beta} l ) } \bigg\{ \cosh \big( \frac{\ell_t l}{\beta L} w\big)  \big[ 1 - \frac{\pi \ell_t l}{\beta L} \coth (\frac{\pi \ell_t l}{\beta L} ) \big] -2 p \frac{\pi \ell_t l}{\beta L}  \sinh ( \frac{\ell_t l \pi }{\beta L } )  \\
&-2 \sinh \Big( \frac{\ell_t l p \pi }{\beta L } \Big)   \sinh \Big( \frac{\ell_t l (w - p\pi)}{\beta L } \Big) \Big[ 1  - \frac{\pi \ell_t l}{\beta L} \coth (\frac{\ell_t l (w - p\pi)}{2L} ) \Big] \bigg\}
\;.  \
\end{split}
\end{align} 
There are two different contributions to the entanglement entropy for $\alpha=0$ in the high temperature limit. The terms in the first line \eqref{HighTEEWithPTerms1} dominate over the rest of the terms as $\beta \to 0$.  

Let us briefly mention the entropies in the presence of chemical potential and current source, $\mu \neq 0$ and $J \neq 0$. The result turns out to be almost the same as \eqref{HighTEEWithPTerms1}. The first line in \eqref{HighTEEWithPTerms1} is the same due to the complex conjugate even in the presence of $\mu,J$. The rest result of  \eqref{HighTEEWithPTerms1} requires a simple multiplicative factor $ \cos (2\pi \mu l)$ inside the $l$ summation. Thus we have a simple $\mu$ dependence. Note that there is no explicit $J$ dependence because $J$ always enter with the combination $\alpha J$, which is removed by setting $\alpha=0$. Nevertheless, the $J$ dependence is hidden in the $a+J=1/2$ condition for the anti-periodic fermions. 

Finally, we discuss the periodic fermions that has $\vartheta_2$ instead $\vartheta_3$. Thus we can compute the high temperature limit of the periodic fermions by studying the S-dual formula for $\vartheta_2$ which is $ \vartheta_2 (z|\tau) = (-i \tau)^{-1/2} e^{-\pi i z^2/\tau} \vartheta_4 (z/\tau|-1/\tau) $. Now, the difference between $\vartheta_3$ and $\vartheta_4$ is a couple of signs. The resulting entanglement entropy can be found by removing the factor $(-1)^{l-1}$ in \eqref{HighTEEWithPTerms1}. \\

\noindent {\it $\square$ Entanglement entropy $S_{n,H}^{w,\mu,J}$ with $\alpha \neq 0$. } 

For the high temperature limit, we can ignore the $\beta$ dependence in $\tau$ for $2\pi \tau = 2\pi \tau_1 + i 2\pi \tau_2 = \alpha+ i \beta \to \alpha$. Then the entropy formula for the anti-periodic fermion goes 
\begin{align} \label{TopologicalEEFormulaJmuHighTAlpha}
	\log \tilde Z^H_{s}[n] &=\!\! \sum_{k=-{(n-1)}/{2}}^{{(n-1)}/{2}} \log \Big| \frac{e^{-\frac{i2\pi^2}{\alpha} (\frac{\ell_t \alpha_{w,k} }{2\pi L} + \frac{\alpha J}{2\pi} )^2} }{e^{-\frac{i2}{\alpha} ( \alpha J)^2}}  \frac{\vartheta_3 (\frac{\ell_t}{L} \frac{\alpha_{w,k}}{\alpha} +J|\frac{\-2\pi}{\alpha})}{\vartheta_3 (J|\frac{\-2\pi}{\alpha})} \Big|^2 \nonumber \\
	&= 2\sum_{m=1}^{\infty} \sum_{k=-{(n-1)}/{2}}^{{(n-1)}/{2}}\log \big( \frac{\cos (2\pi J + \frac{2\pi \ell_t}{\alpha L} [\frac{k}{n}+ \frac{w}{2\pi}+l_k]) + \cos (\frac{4\pi^2}{\alpha} [m-\frac{1}{2}]) }{\cos (2\pi J ) + \cos (\frac{4\pi^2}{\alpha} [m-\frac{1}{2}])} \big) \;, 
\end{align} 
where we expand the theta functions with index $m$ and use $ (1+e^{i a+ i b})(1+e^{i a- i b})(1+e^{-i a+ i b})(1+e^{-i a- i b}) = 4 (\cos a + \cos b)^2$. For this case, we can not do the $k$ sum analytically. We need to rely on some other methods. It will be interesting to look into this further. 

\section{Mutual Information} \label{sec:MutualInformation}

The Mutual information is useful to measure the entanglement between two intervals, $A$ and $B$ of length $\ell_A$ and $\ell_B$ separated by $\ell_C$. The mutual information is finite and free of UV divergences. 

In the presence of the chemical potential and current source, the entropies have been considered to include $\mu$ and $J$ in \cite{Herzog:2013py}\cite{Kim:2017xoh}\cite{Kim:2017ghc}. Here, we generalize this further to include the Wilson loops parameter $w$ that is the same for the sub-systems denoted by $A$ and $B$. Similar to the entropies, the mutual information factories into two different contributions. 
\begin{align}
I^w_n (A,B) &= I^{w,0}_n (A,B) + I^{w,\mu,J}_n (A,B) \;.
\end{align} 
The first contribution is generalized to include the Wilson loops contribution as
\begin{align} \label{GeneralMIW1}
I^{w,0}_n (A,B) &=\frac{1}{1\- n} \Big( \sum_{k=-{(n-1)}/{2}}^{{(n-1)}/{2}} \alpha_{w,k}^2 - \alpha_{w,0}^2 \Big)  \log \Bigg| \frac{\vartheta [\substack{1/2 \\ 1/2 }](\frac{\ell_A+\ell_B+\ell_C}{2\pi L}|\tau) ~\vartheta [\substack{1/2 \\ 1/2 }](\frac{\ell_C}{2\pi L}|\tau)}{\vartheta [\substack{1/2 \\ 1/2 }](\frac{\ell_A+\ell_C}{2\pi L}|\tau) ~\vartheta [\substack{1/2 \\ 1/2 }](\frac{\ell_B+\ell_C}{2\pi L}|\tau)} \Bigg|^{2} \;. 
\end{align} 
Where $\alpha_{w,k}= \frac{k}{n} + \frac{w}{2\pi} + l_k$. This sum for the expression inside the parenthesis has been evaluated in \S \ref{sec:FiniteEEMuJZero} for the general topological transitions with appropriate normalization factors 
\begin{align}
\frac{1}{1- n} \Big( \sum_{k=-(n-1)/2}^{(n-1)/2} \alpha_{w,k}^2 - \alpha_{w,0}^2 \Big)  = \frac{n+1}{12 n} + Q\frac{w}{\pi} - Q^2\frac{1}{n } \;, 
\end{align}
where $Q$ is the number of topological transitions. The rest of the factors have been already studied \cite{Kim:2017xoh}\cite{Kim:2017ghc}, and we will not consider them further. Note that this spin structure independent mutual information depends on the separation distance $\ell_C$ as well as the sub-system sizes $\ell_A$ and $\ell_B$.

The second contribution has the dependences on current source $J$, chemical potential $\mu$ and the Wilson loops parameter $w$. We can write it as $ I^{w,\mu,J}_n  = \frac{1}{1- n}  \big[ \log \tilde Z_{s}^I[n] - n \tilde Z^I_s[1] \big]$
\begin{align} \label{GeneralMIW2}
\tilde Z_{s}^I[n] = \prod_{k=-{(n-1)}/{2}}^{{(n-1)}/{2}}  \Bigg| \frac{\vartheta [\substack{1/2-a-J \\ b-1/2 }]( \frac{\ell_A \alpha_{w,k}}{2\pi L} \+ \tau_1 J \+ i \tau_2 \mu|\tau)~ \vartheta [\substack{1/2-a-J \\ b-1/2 }]( \frac{\ell_B  \alpha_{w,k}}{2\pi L} \+ \tau_1 J \+ i \tau_2 \mu|\tau)}{\vartheta [\substack{1/2-a-J \\ b-1/2 }](\tau_1 J \+ i \tau_2 \mu|\tau)~\vartheta [\substack{1/2-a-J \\ b-1/2 }](\frac{(\ell_A+\ell_B)  \alpha_{w,k}}{2\pi L} \+ \tau_1 J \+ i \tau_2 \mu|\tau)} \Bigg|^2 \;. 
\end{align} 
Here we note that the spin structure dependent entropies are independent of the separation distance $\ell_C$ between the sub-system sizes. We guess that this is due to the total system size is divided up in four different regions, two for the sub-systems $\ell_A$ and $\ell_B$, two for their separations $\ell_C$ and $2\pi L - (\ell_A + \ell_B + \ell_C)$. The two separation distances sum up as $\ell_C + 2\pi L - (\ell_A + \ell_B + \ell_C) = 2\pi L - (\ell_A + \ell_B)$ to be independent of $\ell_C$. This is in contrast to the spin structure independent mutual information.   

Various properties of this mutual information has been studied previously. Let us consider a low temperature limit $\beta \to \infty$ for the anti-periodic fermions with $b=1/2, a+J=1/2$. We also include the $Q$ transitions for the range of the Wilson loop parameter $(2Q-1)\pi /n \leq w < (2Q+1)\pi /n$.
\begin{align} \label{MutualInformationLowTAP}
\log \tilde Z_{s}^I[n] 
&=  \sum_{k=-{(n-1)}/{2}}^{{(n-1)}/{2}} \log  \Big| \frac{\vartheta _3( \frac{\ell_A \alpha_{w,k}}{2\pi L} + \tau_1 J + i \tau_2 \mu|\tau)~ \vartheta_3( \frac{\ell_B  \alpha_{w,k}}{2\pi L} + \tau_1 J + i \tau_2 \mu|\tau)}{\vartheta_3(\tau_1 J + i \tau_2 \mu|\tau)~\vartheta_3(\frac{(\ell_A+\ell_B) \alpha_{w,k}}{2\pi L} + \tau_1 J + i \tau_2 \mu|\tau)} \Big|^2  \;. 
\end{align} 
Upon examining \eqref{MutualInformationLowTAP}, we observe that 
\begin{align} \label{ThetaManipulations}
	&\log  \Big| \frac{\vartheta _3( \ell_A ~ \cdots )~ \vartheta_3( \ell_B ~ \cdots )}{\vartheta_3(0 ~ \cdots )~\vartheta_3((\ell_A+\ell_B) ~ \cdots)} \Big|^2 \nonumber \\
	&= \log  \Big| \frac{\vartheta _3( \ell_A ~ \cdots)}{\vartheta_3(0 ~\cdots)} \Big|^2 
	+ \log  \Big| \frac{\vartheta_3( \ell_B ~ \cdots)}{\vartheta_3(0 ~ \cdots)} \Big|^2 
	-\log  \Big| \frac{\vartheta_3((\ell_A+\ell_B)~ \cdots)}{\vartheta_3(0 ~ \cdots)} \Big|^2  \;. 
\end{align} 
The computations with two theta functions has been carried out in various places in this paper, for example \eqref{LowTEEFiniteSNaJ}. Now, the answer is sum of these expressions similar to \eqref{LowTEEFiniteSNaJ} for three different intervals, $\ell_A$, $\ell_B$ and $\ell_A+\ell_B$, which is lengthy. We do not explicitly write them here. We note that the spin dependent mutual information is independent of the separation $\ell_C$ between the two entanglement regions $\ell_A$ and $\ell_B$. The full answer for the mutual information is the sum \eqref{GeneralMIW1} and \eqref{GeneralMIW2} for the anti-periodic fermions.

We would like to have some further remarks on the mutual information. First, let us consider the zero temperature limit. The explicit computations for two theta functions \eqref{ThetaManipulations} has been done in the section \S \ref{sec:GeneralLowTEEMuJWGan} for the anti-periodic fermions, and the final entanglement expression is given in \eqref{LowTEEFiniteSNaJ} for $-1/2 < \mu <1/2$. One needs care for the range of $\mu$ for the low temperature limit as mentioned previously. When $\mu$ is half an integer, there is a finite contribution at zero temperature limit. These half integer values of $\mu$ have been identified as the energy levels of the Dirac fermion on a circle. Thus entanglement entropy can be useful to identify energy levels of quantum theories \cite{Kim:2017xoh}\cite{Kim:2017ghc}. For other values of $\mu$, refer to \cite{Kim:2017ghc}. We also check that the result reduces to the previous computation, equation (4.5) in \cite{Kim:2017ghc}, when $w=0$. 

The computation for the periodic fermions can be done similarly. The result is the same as the spin structure independent part, while the spin structure dependent part has some modifications, $m-1/2 \to m$ and additional term in the summation over the index $l$. This new contribution is from the $\cos$ factor in the $\vartheta_2$ for the periodic fermions. For more details, refer to \cite{Kim:2017ghc}. For the large radius limit, the spin dependent part $I^{w,\mu,J}_n$ of the mutual information also vanishes, while $I^{w,0}_n (A,B)$ given in \eqref{GeneralMIW1} survives.  \\

$\star$ {\it High temperature limit} \\
Finally, we comment on the high temperature limit $\beta\to 0$. The general formulas \eqref{GeneralMIW1} and \eqref{GeneralMIW2} can be rewritten by using the modular transformations as we did in the previous section. We first consider \eqref{GeneralMIW1} without the dependence on $\mu, J$. 
\begin{align} \label{MutualInformation1WHighT}
\begin{split}
	I^{w,0}_{n,H} (A,B) &= \frac{1}{1\- n} \Big( \sum_{k=-{(n-1)}/{2}}^{{(n-1)}/{2}} \alpha_{w,k}^2 - \alpha_{w,0}^2 \Big)   \\
	& \times \log \bigg. \Big| \frac{e^{-\frac{i\pi}{\tau}(\frac{\ell_A+\ell_B+\ell_C}{2\pi L} )^2-\frac{i\pi}{\tau}(\frac{\ell_C}{2\pi L})^2}}{e^{-\frac{i\pi}{\tau}(\frac{\ell_A+\ell_C}{2\pi L})^2 -\frac{i\pi}{\tau}(\frac{\ell_B+\ell_C}{2\pi L})^2}}
	\frac{\vartheta [\substack{1/2 \\ 1/2 }](\frac{\ell_A+\ell_B+\ell_C}{2\pi L \tau}|\frac{-1}{\tau}) ~\vartheta [\substack{1/2 \\ 1/2 }](\frac{\ell_C}{2\pi L \tau}|\frac{-1}{\tau})}{\vartheta [\substack{1/2 \\ 1/2 }](\frac{\ell_A+\ell_C}{2\pi L \tau}|\frac{-1}{\tau}) ~\vartheta [\substack{1/2 \\ 1/2 }](\frac{\ell_B+\ell_C}{2\pi L \tau}|\frac{-1}{\tau})} \Big|^2 \;. 
\end{split}	
\end{align}  
Here we choose the normalization factor so that there exist $n \to 1$ limit. As mentioned, the high temperature limit is sensitive to $\alpha=2\pi \tau_1$. For $\alpha \neq 0$, we compute it by using $ \frac{1}{\tau}= \frac{2\pi}{\alpha + i \beta}\to \frac{2\pi}{\alpha}$. Thus $I^{w,0}_{n,H} (A,B)$ gives us for $\alpha\neq 0$  
\begin{align} \label{GeneralMIW1HighT}
\begin{split}
	I^{w,0}_{n,H} (\alpha\neq 0)  &=- \big( \frac{n+1}{12 n} + Q\frac{w}{\pi} - Q^2 \frac{1}{n} \big)\bigg\{ 2 \log \big( \frac{\sin (\pi \frac{\ell_A+\ell_B+\ell_C}{L \alpha} ) \sin (\pi \frac{\ell_C}{L \alpha})}{\sin (\pi \frac{\ell_A+\ell_C}{L \alpha} ) \sin (\pi \frac{\ell_B+\ell_C}{L \alpha})} \big) \\
	& + 2 \sum_{m=1}^\infty \log \big(
	\frac{\big[ \cos (2\pi \frac{\ell_A+\ell_B+\ell_C}{L \alpha} ) - \cos (\frac{4\pi^2}{\alpha} m)\big]
	\big[ \cos (2\pi \frac{\ell_C}{L \alpha} ) - \cos (\frac{4\pi^2}{\alpha} m )\big]}{\big[ \cos (2\pi \frac{\ell_A+\ell_C}{L \alpha} ) - \cos (\frac{4\pi^2}{\alpha} m)\big] 
	\big[ \cos (2\pi \frac{\ell_B+\ell_C}{L \alpha} ) - \cos (\frac{4\pi^2}{\alpha} m)\big]} \big) \bigg\}\;.
\end{split}
\end{align} 
We can compute the mutual information for $\alpha=0$ with $\frac{1}{\tau}=-i \frac{2\pi}{\beta} $. There is an additional term due to the change for sine function to the hyperbolic sine in the exponential factor in \eqref{MutualInformation1WHighT}. Thus we get 
\begin{align} \label{MutualInfoWHT1ZeroAlpha}
\begin{split}
I^{w,0}_{n,H} (\alpha= 0)  &=- \big( \frac{n+1}{12 n} + Q\frac{w}{\pi} - Q^2 \frac{1}{n} \big) \bigg\{  - \frac{2}{\beta} \frac{\ell_A \ell_B}{L^2}  + 
2 \log \big( \frac{\sinh (\pi \frac{\ell_A+\ell_B+\ell_C}{L \beta} ) \sinh (\pi \frac{\ell_C}{L\beta})}{\sinh (\pi \frac{\ell_A+\ell_C}{L \beta} ) \sinh (\pi \frac{\ell_B+\ell_C}{L\beta})} \big)  \\
&+ 2 \sum_{m=1}^\infty \log\big(
\frac{\big[ \cosh (2\pi \frac{\ell_A+\ell_B+\ell_C}{L \beta} ) \- \cosh (\frac{4\pi^2}{\beta}m )\big]
\big[ \cosh (2\pi \frac{\ell_C}{L \beta} ) \- \cosh (\frac{4\pi^2}{\beta}m )\big]}{\big[ \cosh (2\pi \frac{\ell_A+\ell_C}{L \beta} ) \- \cosh (\frac{4\pi^2}{\beta} m)\big] 
\big[ \cosh (2\pi \frac{\ell_B+\ell_C}{L \beta} ) - \cos (\frac{4\pi^2}{\beta} m)\big]} \big)  \bigg\} \;. 
\end{split}
\end{align} 
These results for the case $w=0$ have been reported previously \cite{Kim:2017ghc}. Note that the spin structure independent contributions depend on the sub-system sizes, $\ell_A, \ell_B$ and their separation $\ell_C$.

The chemical potential and current source dependent mutual information has further contributions. We consider $I^{w,\mu,J}_{n,H}  = \frac{1}{1 - n}  \left( \tilde I_{H} [n]  - n \tilde I_{H}[1] \right)$. 
\begin{align} \label{GeneralMIW2HighT}
\begin{split}
\tilde I_{H} [n]
&=\sum_{k=-{(n-1)}/{2}}^{{(n-1)}/{2}} \log  \Big| \frac{e^{-\frac{i\pi}{\tau}(\alpha_{w,k} \frac{\ell_A}{2\pi L} + \tau_1 J + i \tau_2 \mu)^2}~ e^{-\frac{i\pi}{\tau}(\alpha_{w,k} \frac{\ell_B}{2\pi L} + \tau_1 J + i \tau_2 \mu)^2}}{e^{-\frac{i\pi}{\tau}(\tau_1 J + i \tau_2 \mu)^2}~ e^{-\frac{i\pi}{\tau}(\alpha_{w,k} \frac{\ell_A+\ell_B}{2\pi L} + \tau_1 J + i \tau_2 \mu)^2}}  \\
&\qquad \times \frac{\vartheta [\substack{1/2-b \\ 1/2-a-J }](\alpha_{w,k} \frac{\ell_A}{2\pi L \tau} + \frac{\tau_1 J + i \tau_2 \mu}{\tau}|\frac{-1}{\tau})~ \vartheta [\substack{1/2-b \\ 1/2-a-J  }](\alpha_{w,k} \frac{\ell_B}{2\pi L \tau} + \frac{\tau_1 J + i \tau_2 \mu}{\tau}|\frac{-1}{\tau})}{\vartheta [\substack{1/2-b \\ 1/2-a-J  }](\frac{\tau_1 J + i \tau_2 \mu}{\tau}|\frac{-1}{\tau})~\vartheta [\substack{1/2-b \\ 1/2-a-J  }](\alpha_{w,k} \frac{\ell_A+\ell_B}{2\pi L \tau} + \frac{\tau_1 J + i \tau_2 \mu}{\tau}|\frac{-1}{\tau})} \Big|^2 \;. \end{split}
\end{align} 
As mentioned the spin structure dependent entropies are independent of the separation distance $\ell_C$ between the two sub-systems, $A$ and $B$. The four exponential factors yield the value $1$ for $\alpha \neq 0$ and $ \beta \to 0$ because they have only the imaginary exponents. For $\alpha=0$, $\beta$ is important even though we take $\beta\to 0$ limit. The exponential factors contribute to the mutual information $I^{w,\mu,J}_{n \to 1,H}$ in the $n\to 1$ limit for the $Q$ topological transitions as 
\begin{align}
	-  \big( \frac{n+1}{12 n} + Q\frac{w}{\pi} - Q^2 \frac{1}{n} \big) \frac{2}{\beta} \frac{\ell_A \ell_B}{L^2} \;, \qquad \text{for} \quad \alpha=0 \;.
\end{align}

For the other four Theta functions in \eqref{GeneralMIW2HighT}, we consider the anti-periodic fermions $a+J = b=1/2$ with $\alpha=0, \beta\to 0$. We have four different $\theta_3$ that can be read off by using the trick \eqref{ThetaManipulations}. Thus we have the final result that has three different sets of the result \eqref{HighTEEWithPTerms1} with $\ell_t$ replaced by $\ell_A$, $\ell_B$ and $\ell_A+\ell_B$ (with relative - sign for the latter). The result for $\alpha\neq 0$ is similar to \eqref{TopologicalEEFormulaJmuHighTAlpha} with two additional factors with different $\ell$'s. Periodic fermions have similar results with the modifications $m-1/2 \to m$. With the detailed result, \eqref{HighTEEWithPTerms1}, given in the previous sections, one can read off the results straightforwardly. 

\section{Conclusion} \label{sec:Conclusion}

In this paper, we generalize previously available results on the R\'enyi and entanglement entropies of the 2 dimensional Dirac fermions in the presence of gauge fields. First, we systematically extend the entropies to have the general formulas in the presence of the Wilson loops as well as the chemical potential and current source by exhausting all possible components of the electromagnetic twist vertex operators in $\mathbb{Z}_n$ orbifold theories on torus in \S \ref{sec:Setup}. These possible components include the electromagnetic parameters $w$ and $k/n$ ($k= -(n-1)/2, -(n-3)/2, \cdots, (n-1)/2$) in addition to the chemical potential $\mu$, current source $J$ and the twist boundary conditions on the two cycles on torus. The case without the topological Wilson loops was previously considered in \cite{Kim:2017xoh}\cite{Kim:2017ghc}. 

Second, we generalize the results of \cite{Belin:2013uta} by including the general Wilson loop winding sector with a parameter $\gamma$, $-2\gamma \pi \leq w < 2\gamma \pi$, as well as by using the normalization factors $\tilde \alpha_w$ depending on the Wilson loops parameter by choosing the following requirements \\ \vspace{-0.2in} 

\indent	\qquad I. R\'enyi entropy has a smooth entanglement entropy limit for $n \to 1$. \\
\indent	\qquad II. R\'enyi and entanglement entropies are continuous across different topological sectors. \\ \vspace{-0.2in} 

\noindent These two conditions can be met with the following choice. 
\begin{align} 
\alpha_{w,0}^2 \equiv \lim_{n \to 1} \Big( \sum_{k=-{(n-1)}/{2}}^{{(n-1)}/{2}} \alpha_{w,k}^2 \Big) + \tilde \alpha_w (n-1)\;.
\end{align}	
The first part ensures that the R\'enyi entropy has a smooth limit for entanglement entropy, while the second part can be used to make the R\'enyi and entanglement entropies be continuous. 

We compute the entropies with the general R\'enyi and entanglement entropies. We list some salient features of our new findings here. 
\begin{itemize}
	\item Infinite space entanglement entropy: The spin structure independent R\'enyi and entanglement entropies have non-trivial dependences on the Wilson loops parameter $w$. In particular, in the infinite space limit, entanglement entropy at the $p$-th topological sector is given by \eqref{EEIndSpinSWithPEELargeLimit}. 
	\begin{align} 
	S_{n\to 1,p}^{w,0} = 2 \bigg(  \frac{Qw}{\pi} - \frac{6Q^2 -1}{6} \bigg) \log |u-v|  \;, 
	\end{align}
	where $|u-v|$ is the length of the subsystem. The formula works for the topological sector parameterized by a number $Q$ for $(2Q-1)\pi \leq w < (2Q+1)\pi$. The quantitative features are depicted in the figure \ref{fig:EESpinIndependent} across the topological sectors. We necessarily consider the entire real range to include all the possible winding numbers of $w$. If one considers a specific winding sector $\gamma$, then she/he can restricts the range for $w$ as $-2\gamma \pi \leq w < 2 \gamma \pi $. In figure \ref{fig:EESpinIndepPEELimit}, we depict the R\'enyi and entanglement entropies for $n=1,2,3,4,5$ so that one can see how the entanglement entropy limit is approached.  
	
	This result turns out to be exact in infinite space, because the spin structure dependent parts do not contribute to the infinite space limit. Note that chemical potential $\mu$ and current source $J$ do not contribute to the infinite space limit as demonstrated here and also in \cite{Kim:2017xoh}\cite{Kim:2017ghc}. Previously, it was claimed that entanglement entropy is independent of the chemical potential in infinite space based on a more general field theory arguments based on symmetries in \cite{CardySlides:2016}, where topological sectors were not considered. Thus we confirm the result in \cite{CardySlides:2016}. 
	
	\item Finite size (space) effects: The spin structure dependent R\'enyi and entanglement entropies have also non-trivial dependence on the Wilson loops parameter $w$ as in \eqref{LowTEEFiniteSNa} for $\mu=J=0$ as well as in \eqref{LowTEEFiniteSNaJ} for $\mu\neq 0$ and $J\neq 0$. The quantitative features for the $n=2$ and $n=3$ R\'enyi entropies and entanglement entropy have been depicted in the figure \ref{fig:RenyiLowT}, while the sub-system size dependences of entanglement entropy in the figure \ref{fig:EELowT}. 
	
	- When we introduce the chemical potential dependence $\mu$, the entanglement entropy has non-trivial dependences on the chemical potential when the latter coincides with one of the energy levels of a quantum system at zero temperature. This feature do not interfere with the effects of the Wilson loops. This has been fully investigated in \cite{Kim:2017xoh}\cite{Kim:2017ghc}. 
	
	- When we introduce the dependence on current source $J$, the entropies reveal oscillating behavior as a function of $J$ with a periodicity $l \alpha J=2\pi$, where $l$ is an expansion parameter, and $\alpha = 2\pi \tau_1$ one of the moduli parameter of torus. The typical dependence is depicted in \ref{fig:EELowTGeneralJ}. 
	
	- These spin structure dependent entropies $S_n^{w,\mu,J}$ vanish as fast as $ \mathcal O(\ell_t^2/L^2)$ as $\ell_t/L \to 0 $.
\end{itemize}

Our new results are based on systematic analysis based on the results of the well established $\mathbb{Z}_n$ orbifold conformal field theory. We include all the possible elements of background gauge fields. It is interesting to verify these findings in the real world. \\ 

%\section*
\noindent {\bf Acknowledgement:} 
We are grateful to Sumit Das, Andrea Erdas and Moshe Goldstein for numerous discussions and valuable comments. \\ 

\end{document}